\documentclass[%
 aip,
 amsmath,amssymb,
 reprint,%
]{revtex4-1}

\usepackage{graphicx}% Include figure files
\usepackage{dcolumn}% Align table columns on decimal point
\usepackage{bm}% bold math
\usepackage[utf8]{inputenc}
\usepackage[T1]{fontenc}
\usepackage{mathptmx}
\usepackage{xcolor}
\usepackage{hyperref}
\usepackage{float}
\newcommand{\RomanNumeralCaps}[1]
{\MakeUppercase{\romannumeral #1}}

\begin{document}

\preprint{AIP/123-QED}

\title[Blob interaction in 2D scrape off layer simulations]{Blob interaction in 2D scrape off layer simulations}

\author{G.~Decristoforo}
	\email{gregor.decristoforo@uit.no (corresponding author)}
	\altaffiliation{Department of Physics and Technology, UiT The Arctic University of Norway, NO-9037 Troms{\o}, Norway}
\author{F.~Militello}
	\altaffiliation{CCFE, Culham Science Centre, Abingdon OX14 3DB, United Kingdom}
\author{O.~E.~Garcia}
	\altaffiliation{Department of Physics and Technology, UiT The Arctic University of Norway, NO-9037 Troms{\o}, Norway}
\author{T.~Nicholas}
	\altaffiliation{York Plasma Institute, Department of Physics, University of York, Heslington, York YO10 5DD, United
Kingdom}
\author{J.~Omotani}
	\altaffiliation{CCFE, Culham Science Centre, Abingdon OX14 3DB, United Kingdom}
\author{C.~Marsden}
	\altaffiliation{University of Birmingham, School of Physics and Astronomy, Edgbaston, Edgbaston Park Road,	Birmingham	B15 2TT, United Kingdom}
\author{N.~Walkden}
	\altaffiliation{CCFE, Culham Science Centre, Abingdon OX14 3DB, United Kingdom}

\date{\today}

\begin{abstract}
Interaction of coherent structures known as blobs in the scrape-off layer of magnetic confinement fusion devices is investigated. Isolated and interacting seeded blobs as well as full plasma turbulence are studied with a two dimensional fluid code. The features of the blobs (size, amplitude, position) are determined with a blob tracking algorithm, which identifies them as coherent structures above a chosen density threshold and compared to a conventional center of mass approach. The agreement of these two methods is shown to be affected by the parameters of the blob tracking algorithm. The benchmarked approach is then extended to a population of interacting plasma blobs with statistically distributed amplitudes, sizes and initial positions for different levels of intermittency. As expected, for decreasing intermittency, we observe an increasing number of blobs deviating from size-velocity scaling laws of perfectly isolated blobs. This is found to be caused by the interaction of blobs with the electrostatic potential of one another, leading to higher average blob velocities. The degree of variation from the picture of perfectly isolated blobs is quantified as a function of the average waiting time of the seeded blobs.
\end{abstract}

\maketitle

\section{Introduction}\label{intro}

In tokamaks and other magnetically confined plasma experiments, particle transport in the plasma edge region is dominated by turbulence-driven coherent structures of high density and temperature called blobs or filaments. This can lead to large erosion on the reactor walls and can contribute to the power loads to divertor targets \cite{antar2001experimental,antar2003universality,kirk2006filament,dudson2008experiments,ayed2009inter}. These structures have been observed in multiple plasma devices in all operation regimes using reciprocating or wall mounted Langmuir probes \cite{boedo2001transport,rudakov2002fluctuation,boedo2003transport,garcia2007fluctuations,garcia2007collisionality,militello2013experimental}, fast visual cameras \cite{kirk2006filament,dudson2008experiments, ayed2009inter, farley2017analysis, walkden2017quiescence, walkden2017identification, walkden2018fluctuation} and gas puff imaging \cite{zweben2002edge,terry2003observations, myra2006blob, grulke2006radially, zweben2015edge, zweben2016blob}.\\
In addition to experimental evidence, theoretical understanding of the underlying physical mechanism of blob propagation has been developed in the last 20 years \cite{d2004blob, d2011convective, krasheninnikov2008recent, garcia2009blob}. It is understood that the basic mechanism responsible for the radial transport of blobs arises due to grad-B and curvature drifts leading to a charge polarization in the plasma blob/filament. The resulting electric field gives rise to an \textbf{E}$\times$\textbf{B} drift that propels the blob across the magnetic field. Since detailed physical models increase the analytical complexity significantly, the scientific community relies on numerical simulations of isolated blobs and fully turbulent simulations of the scrape off layer. Numerical simulations in two dimensions \cite{garcia2004computations, russell2009saturation, militello2012simulations, militello2013numerical, kube2011velocity, kube2016amplitude, wiesenberger2017unified} and three dimensions \cite{angus2012effect, walkden2013characterization, ricci2015approaching, tamain20143d, easy2014three, easy2016investigation, militello2016multi, riva2016blob, riva2019three} have enhanced the understanding of the underlying mechanisms of blob and filament propagation in the scrape off layer. \\
Most of these numerical simulations investigate idealized isolated blobs modeled as positive symmetrical Gaussian perturbations on a constant plasma background. This approach has provided an effective way of investigating the influence of specific physical effects, such as finite Larmor radius effects \cite{wiesenberger2014radial}, electromagnetic effects \cite{lee2015electromagnetic} or parallel electron dynamics \cite{angus2012effects} on the blob velocity, coherence and lifetime. Scaling laws describing the radial blob velocity depending on it's amplitudes and size \cite{kube2011velocity,kube2016amplitude} have been developed, and different regimes determined by various physical parameters have been discovered \cite{myra2006collisionality,russell2007collisionality}. \\
Despite this progress, understanding how well these scaling laws describe blobs in fully turbulent scenarios where they interact with each other is non-trivial. Previous work has shown that single blobs in close proximity do interact through the electric potential they generate\cite{militello2017interaction}. This analysis was performed on two spatially separated seeded blobs on a constant plasma background, and therefore does not address the complexity of a fully turbulent environment. In our work, we expand the investigation by starting from isolated blob simulations and then extending our analysis to decreasingly intermittent systems, until we consider fully turbulent scrape-off layer plasma. To bridge these two extremes we use a stochastic model of multiple randomly seeded blobs where blob amplitudes, widths, initial positions and the waiting times between consecutive blobs are randomly sampled from distribution functions. \\
In order to track blobs in these intermittent and turbulent scenarios, we developed a new tool that allowed us to go beyond what was done in the past. Our blob tracking algorithm provides specific parameters such as trajectory, velocity, size and amplitude over the lifetime of specific blobs. Tracking algorithms using either simple threshold methods, defining every coherent structure above a chosen density threshold as a blob, or convolutional neural networks have been presented and applied on two and three dimensional data \cite{nespoli2017blob, nespoli20193d, paruta2019blob}. For our analysis we choose the threshold method, since it provides a simple and consistent definition for blobs in the isolated and fully turbulent case and is easy to implement. Applying blob tracking techniques on experimental measurements on high speed imaging data using i.e. a watershed algorithm \cite{farley2017analysis} is complicated by the spatial and temporal resolution of the measurement techniques. This algorithm is based on fitting two dimensional Gaussians to local density maxima in order to
extract the position, widths, amplitudes and positions of the fluctuations.
\\
The structure of this publication is as follows: In section \RomanNumeralCaps{2} we present the equations of the physical model that we use for our further analysis. In section \RomanNumeralCaps{3} we present a detailed description of the implementation of the blob tracking algorithm and discuss all relevant parameters of this method. Furthermore, we apply this algorithm on isolated seeded blob simulations in section \RomanNumeralCaps{4}  and compare the results to a conventional center of mass approach. In section \RomanNumeralCaps{5} we extend this analysis on a model seeding multiple blobs randomly. We start with the case of identical amplitudes and starting positions for different intermittency parameters, extend this analysis to random initial positions and finally to a model including random blob amplitudes. In all cases we compare the measurements to the isolated blob simulations. In section \RomanNumeralCaps{6} we finally apply the blob tracking algorithm on fully turbulent scrape off layer simulations and discuss the results in comparison to the previous models.

\section{Physical model}

For our analysis we choose a standard two dimensional (2D), two-field fluid model derived from the Braginskii fluid equations. We assume a quasi-neutral plasma, negligible electron inertia, isothermal electrons, $T_e = \textrm{constant}$, and cold ions, $T_i = 0$. Note that these assumptions for the electron and ion temperatures are taken for the sake of simplification, as experimental measurements of scrape off layer plasmas often show high variations of $T_e$ and $T_i > T_e$ \cite{kovcan2007ion,kocan2012ion,garcia2005interchange}. Nevertheless, this simplified model still captures the fundamental dynamics of the blobs and is therefore sufficient to study their interaction while keeping the number of free parameters of the model relatively low. \\
For our simulations, we use a simple slab geometry to model the plasma evolution perpendicular to the magnetic field, with $x$ and $y$ referring to the radial and the binormal/poloidal direction. The normalized 2D electron particle continuity equation and vorticity equation take the form: 

\begin{equation}
\begin{split}
\frac{d n}{d t} + g \left(\frac{\partial n}{\partial y} -n\frac{\partial \phi}{\partial y} \right) = & D_\perp \nabla_\perp^2 n + S_n - \lambda n e^{-\phi},  
\end{split}
\end{equation}
\begin{equation}
\begin{split}
\frac{d \nabla_\perp^2 \phi}{d t} + \frac{g}{n}\frac{\partial n}{\partial y} =  & \nu_\perp \nabla_\perp^4 \phi +  \lambda \left(1-e^{-\phi}\right),  \\
\end{split}
\end{equation}  

\noindent where $n$ represents the plasma density, $\phi$ the electric potential, $g$ effective gravity,  i.e. interchange drive from magnetic curvature, $S_n$ the plasma source term and $D_\perp$ and $\nu_\perp$ the collisional dissipative terms representing particle diffusivity and viscosity.  The parameter, $\lambda$ is the parallel loss rate of the system. Note that the plasma source term $S_n$ only appears for turbulence simulations and not for seeded blob simulations. The standard Bohm normalization is used for this model equivalent to \cite{easy2014three,easy2016investigation} and is not discussed here for sake of brevity. In addition, we choose $d/dt = \partial / \partial t + \textbf{V}_\textbf{E}\cdot \nabla_\perp$ where $\textbf{V}_\textbf{E} = - \nabla_\perp \phi\times \textbf{B}/B^2$ stands for the $E\times B$ drift. The last term on the right hand side of both the continuity and electron drift vorticity equation results from modelling the parallel losses to the target. \\
The numerical model is implemented in the STORM code \cite{militello2017interaction} which is based on BOUT++ \cite{dudson2009bout++,dudson_ben_2019_3518905}. The code uses a finite difference scheme in the $x$-direction and a spectral scheme in the $y$-direction, time integration is performed by the PVODE solver \cite{byrne1999pvode}. We choose $L_x = 150$ and $L_y = 100$ with a resolution of 256 $\times$ 256 grid points for all runs. The coefficients are representative of a medium sized machine with $g = 1.7\times 10^{-3}$ and $\lambda = 1.8 \times 10^{-4}$. For single isolated blob simulations in section \RomanNumeralCaps{4} we choose $D_\perp = \nu_\perp = 2 \times 10^{-2}$, while for the remaining simulations of section \RomanNumeralCaps{5} and \RomanNumeralCaps{6} $D_\perp = \nu_\perp = 5 \times 10^{-3}$. We choose higher diffusion coefficients for isolated blob simulations in section  \RomanNumeralCaps{4} since the blob coherence stays higher for higher diffusion coefficients.
The source term for the turbulence simulations is
\begin{equation}\label{source}
S_n={\frac {1}{\sigma {\sqrt {2\pi }}}}e^{-{\frac {1}{2}}\left({\frac {x-\mu }{\sigma }}\right)^{2}}
\end{equation}
with $\sigma = 7$ and $\mu = 30$. The source term represents the cross field transport from the core region, but its magnitude and shape here are arbitrary, although convenient. We choose periodic boundary conditions in the $y$-direction and zero gradient boundary conditions in the radial direction for both the density and vorticity fields. For the plasma potential we choose fixed boundary conditions at the radial boundaries $\phi(x = 0) = \phi(x = 150) = 0$.

\section{Numerical implementation of blob tracking}\label{implementation}

\begin{figure*}[ht]
	\centering
	\includegraphics[width=0.4\linewidth]{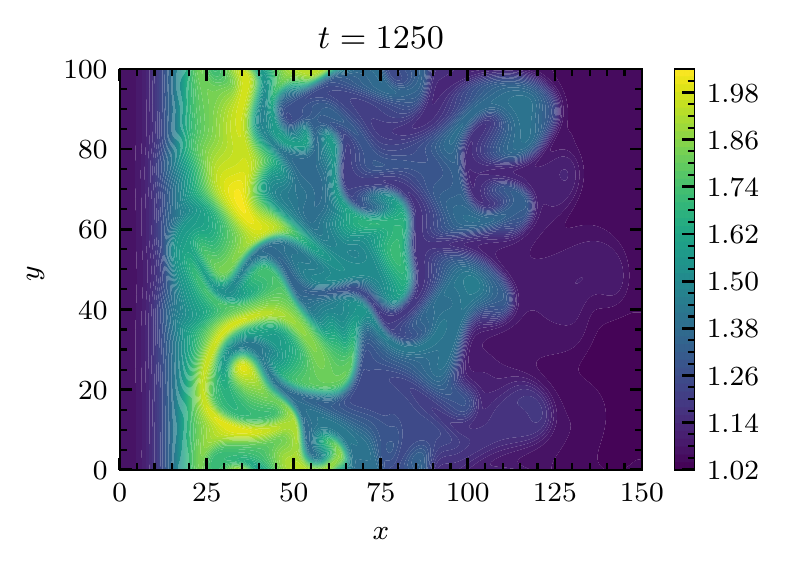}\hfil
	\includegraphics[width=0.4\linewidth]{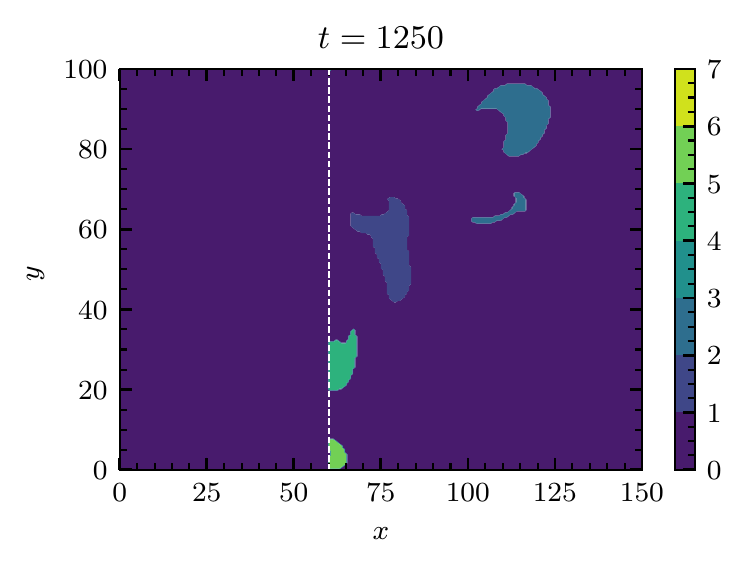}\par\medskip
	\includegraphics[width=0.4\linewidth]{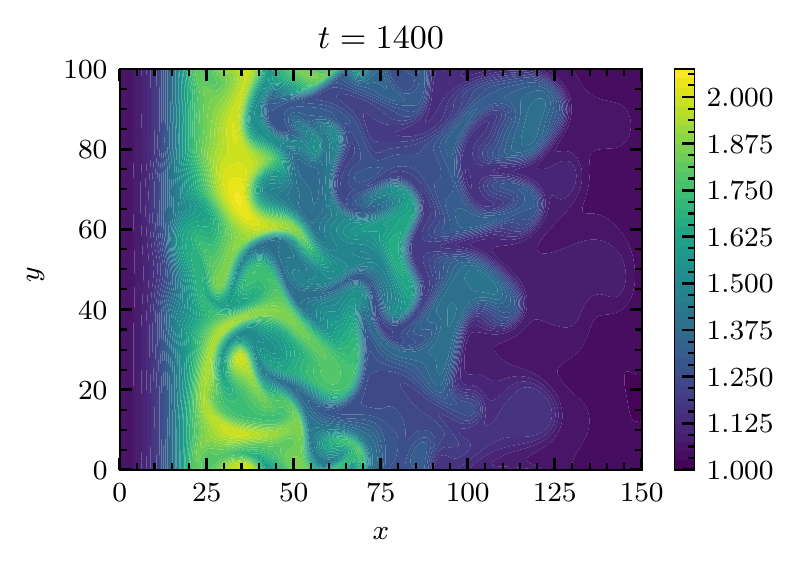}\hfil
	\includegraphics[width=0.4\linewidth]{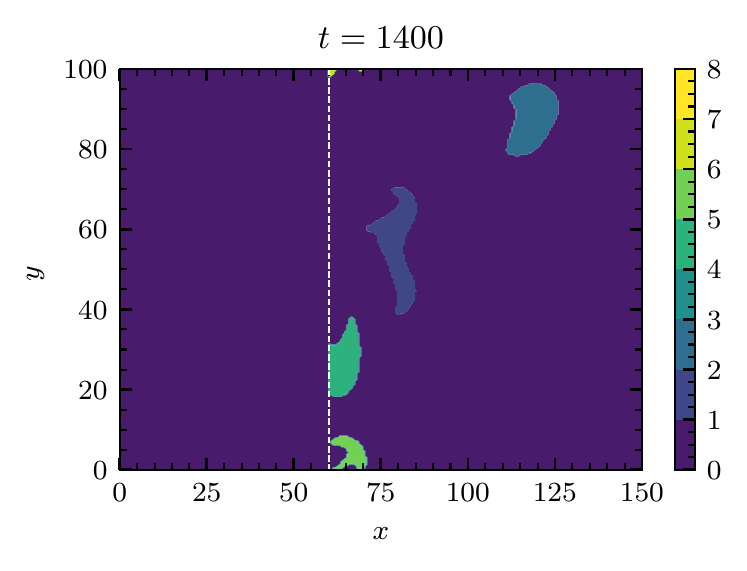}\par\medskip
	\includegraphics[width=0.4\linewidth]{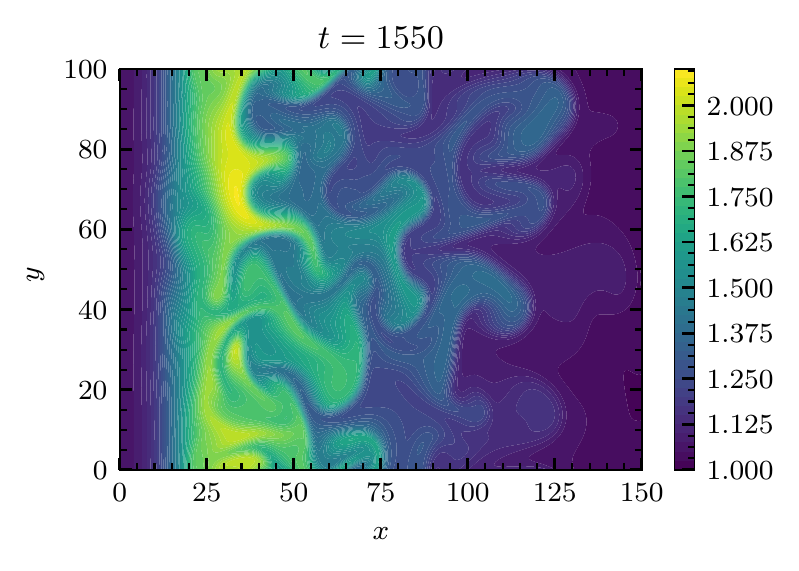}\hfil
	\includegraphics[width=0.4\linewidth]{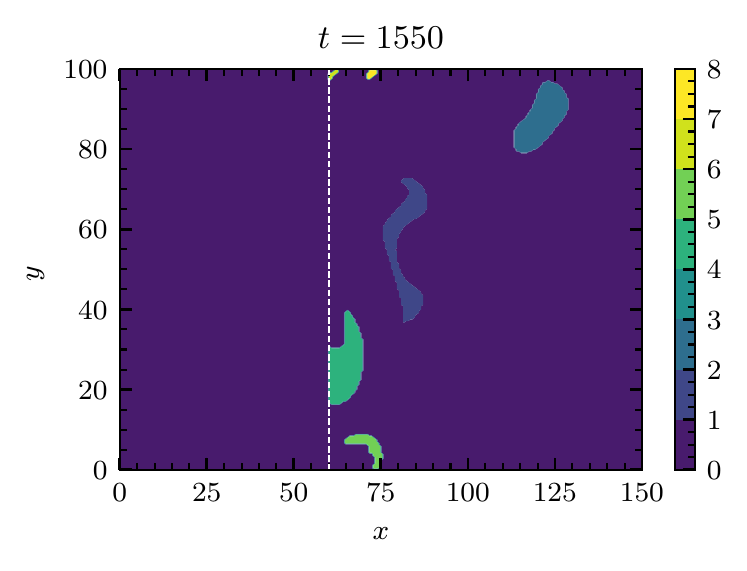}
	\caption{Snapshots of plasma density $n$ and associated blob labels from three closely spaced frames of a turbulence simulation with parameters equivalent to section \RomanNumeralCaps{2}. $x$ refers to the radial and $y$ to the poloidal/binormal coordinate. The colorbar on the right represents the labels of individual detected blobs. The source term on the left side of the domain is excluded from the blob detection algorithm. Radial blob propagation and dissipation is shown for individual detected blobs.}
	\label{tracking_example}
\end{figure*}

The blob tracking algorithm is implemented in Python, employing the xarray library \cite{hoyer2017xarray}. Blobs are identified as positive fluctuations above a certain density threshold. The optimal choice of the thresholding technique depends on the problem at hand, and in the case of the isolated blob simulations presented in Section \RomanNumeralCaps{4}, we take a constant threshold across the whole domain. For the subsequent sections, however, we set a constant threshold on the density field, defined as the total $n$ minus the $y$- and time averaged profile, as this method is more robust for turbulence simulations due to the non-flat average radial profile.
\\
We label the resulting coherent regions using the multi-dimensional image processing library \texttt{scipy.ndimage}. Note that this implementation requires a relatively high temporal resolution of the output files since a blob is only labeled as one coherent structure over time, if the blob spatially overlaps with itself in the next frame. The downside of this approach is the resulting large output files, which slows down the memory bound blob tracking algorithm. In addition, one has to consider the periodic boundary condition in the $y$-direction, since the algorithm will label a blob traveling through the $y$-boundary of the domain as two different objects. For turbulence simulations in section \RomanNumeralCaps{6}, the blob tracking algorithm is only applied in the domain region where $x> 0.4\times L_x$, since we do not include the source term in our analysis. Remember, that the source term is not derived from physical quantities but serves as an artificial numerical term and would heavily interfere the labeling algorithm if included. From these labeled blobs it is straight forward to determine the center of mass of each blob at each time step, its trajectory, radial and poloidal velocity, its amplitude, mass and size over time and its life time. We will use some of these blob parameters for our statistical analysis for different models. We can then apply this method with the identical blob tracking parameters on isolated blobs, statistically seeded blobs and fully turbulent scrape off layer simulations in order to investigate how blob interaction is affected by the plasma intermittency, and its effect on the blob parameters.\\
An example of the blob tracking and labeling methods applied on a turbulence simulation is shown in figure \ref{tracking_example}. This figure shows the plasma density and the associated blobs detected by our algorithm for three different closely spaced frames. The blob tracking further demonstrates in figure \ref{tracking_example}, how individually, detected blobs propagate radially outwards and dissipate over time.  

\section{Isolated seeded blob simulations}

\begin{figure}[H]
	\centering
	\includegraphics[width=8cm]{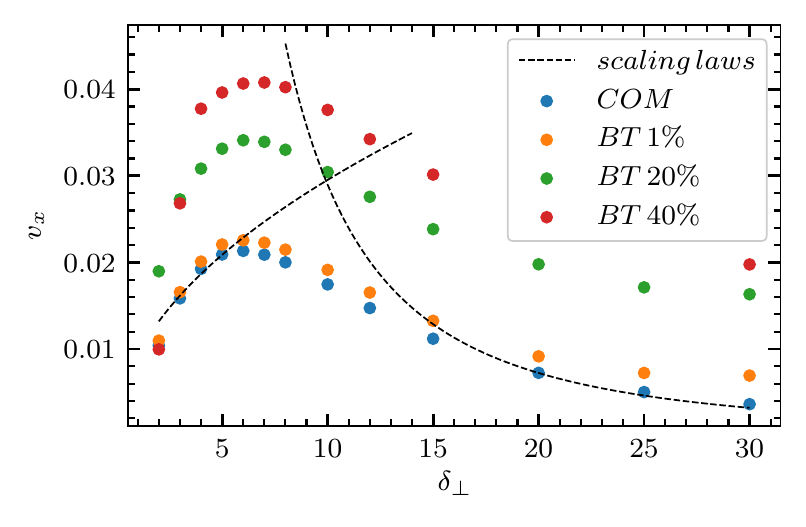}
	\caption{ The dependence of the maximum radial velocity of isolated seeded blobs on their widths compared to theoretical scaling laws. The blue dots refer to the center of mass approach, the other dots to the blob tracking algorithm uses different percentages of the initial amplitude of the blob as a threshold. The radial velocity and the blob width is expressed in normalized units. }
	\label{delta_scan}
\end{figure}

\begin{figure}[t]
	\centering
	\includegraphics[width=8cm]{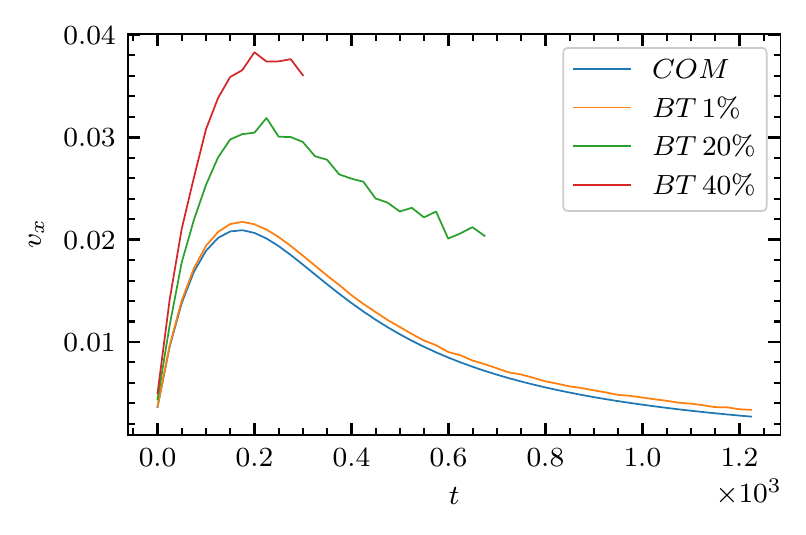}
	\caption{ Radial velocity of an isolated seeded blob width $\delta_\perp = 5$. The blue line refers to the center of mass approach. The other lines refer to the blob tracking algorithm using different percentages of its initial amplitude as the threshold. Radial velocity and time is expressed in normalized units.}
	\label{v-t}
\end{figure}

We begin the analysis by tracking single isolated blobs, seeded on a constant plasma background. We seed a single blob as a symmetrical Gaussian function with amplitude $A$ and width $\delta$ at the initial position, $x_0 = 0.25 \times L_x$ and $y_0 = 0.5 \times L_y$. The blob amplitude is set to be as large as the plasma background, in this case $A = 1$.  We perform a parameter scan from $\delta = 2$ to $\delta = 30$ for the blob width. The blob radial velocity is initially determined by subtracting the plasma background and using a center of mass approach for the whole domain, in order to determine a reference which we use to evaluate our implementation of the blob tracking algorithm. The $x$-component of center of mass of the single blob is therefore calculated by 
\begin{equation}
x_{cm}\left(t\right) = \frac{\int dy \int x\,\left(n\left( x,y,t \right) - n_b\right)\,dx}{\int dy \int \left(n\left( x,y,t \right) - n_b\right)dx}
\end{equation}
\noindent where $n$ stands for the evolving plasma density and $n_b$ for the plasma background density. The $y$-component is calculated analogously and the velocity is determined by a finite difference scheme in time. Next, we determine the blob velocity by using the blob tracking algorithm for three different thresholds. Our algorithm determines the radial velocity of the blob by calculating the center of mass of the plasma region where the plasma density exceeds the threshold and calculates the velocity again by a finite difference scheme in time.\\ 
The results of this analysis are shown in figure \ref{delta_scan}. For all different methods of velocity measurements, we see that the size-velocity dependence follows the theoretical scaling laws studied in previous work \cite{kube2011velocity,myra2006collisionality}. These measurements show that the calculated blob velocity is strongly dependent on the threshold applied for the tracking. For a blob threshold of only one percent of its initial amplitude, we observe that the measured velocity remains very close to the center of mass approach for all widths. This is not surprising, since these two implementations are almost identical for low tracking thresholds. For higher blob thresholds, it is shown that the determined maximum radial velocity increases significantly, as the measured radial velocity for a threshold of $40$ percent of the initial blob amplitude more than doubles the center of mass results. This can be explained by the fact that for high thresholds the algorithm only detects the densest parts of the blob, that tends to propagate faster radially than their less dense regions. This has to be taken into account for further work when applying the blob tracking algorithm on more complex models than singular seeded blob simulations.\\
We further investigate how the blob velocity evolves over the lifetime and how the results change for the different methods. The results of this analysis are shown in figure \ref{v-t} for a relatively small blob width of $\delta = 5$. We observe the absolute velocity dependence on the choice of the threshold of the algorithm. In addition, it is shown that the detected lifetime of the blob for a higher threshold is lower. This can be simply explained by the fact that a narrower blob dissipates energy faster and its amplitude therefore falls under the threshold of the tracking algorithm. The precision of the blob tracking measurement also decreases with higher blob thresholds and smaller blobs. Intuitively, the blob tracking algorithm shows the best performance for wide blobs and low blob thresholds. Due to the good agreement between the results of the center of mass approach and the blob tracking algorithm, we conclude that these methods are consistent, which motivates extending our analysis to  more complex models.

\section{Randomly seeded blob simulations}

\begin{figure}[t]
	\centering
	\includegraphics[width=8cm]{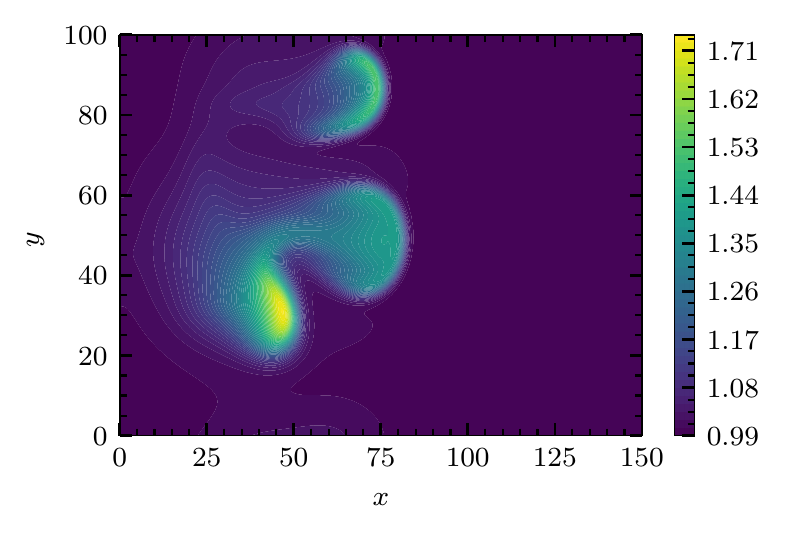}
	\caption{Snapshot of plasma $n$ of a simulation of randomly seeded blobs with different amplitudes. $x$ refers to the radial and $y$ to the poloidal/binormal coordinate. The blob at approx. $y = 90$ propagates radially outwards without interfering with other blobs. At approx. $y=40$ we see two blobs merging into one coherent structure.}
	\label{snapshot}
\end{figure}

\begin{figure}[t]
	\centering
	\includegraphics[width=8cm]{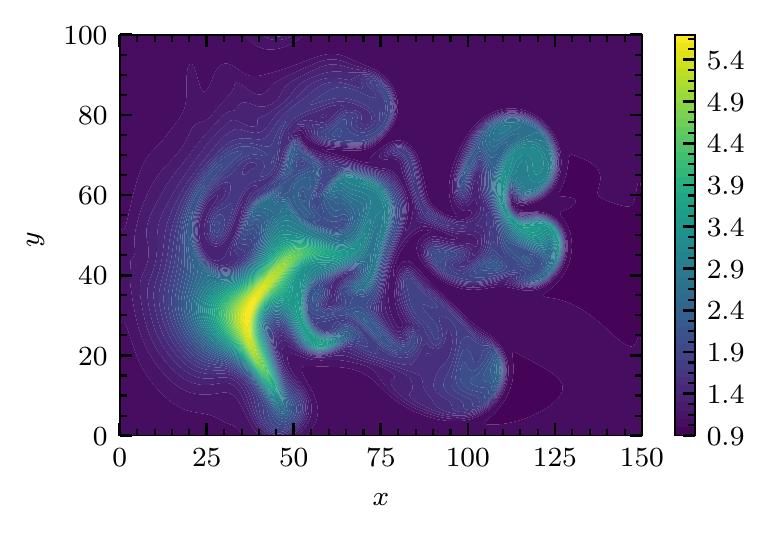}
	\caption{ Snapshot of plasma $n$ of a simulation of randomly seeded blobs with different amplitudes and low intermittency parameter. $x$ refers to the radial and $y$ to the poloidal/binormal coordinate. We observe strong interactions between individual seeded blobs similar to turbulence simulations.}
	\label{snapshot_turbulent}
\end{figure}

The next step of our studies is a more complex model, in which blobs are seeded with random parameters, in particular amplitude, width, initial poloidal/binormal launch position and waiting time between the launch of two consecutive blobs. This model is still artificial but provides valuable insight in blob interaction in a controlled environment. We start our analysis by only keeping waiting times and widths as free parameters and then gradually adding the remaining free parameters to the model. In the most complex case we sample the waiting times and amplitudes from an exponential distribution and the initial poloidal/binormal starting positions and the widths from a uniform distribution. Note, that we choose a uniform distribution for the widths for illustration, even though a log-normal or an exponential distribution would be physically more accurate. Since we intend to compare the velocity-size dependency of detected blobs in this model to isolated blob studies, we choose to sample from a uniform distribution for the sizes to increase the number of big blobs. A snapshot of an example run of this model is shown in figure \ref{snapshot} showing the density field of four seeded blobs with different widths and amplitudes. The blob at approximately $y=90$ propagates in an almost perfectly isolated way radially outwards. The two blobs at approximately $y=50$ show a strong interaction between each other and merge eventually into one coherent structure. A less intermittent case is shown in figure \ref{snapshot_turbulent} where individual blobs interact strongly with each other, resulting in a turbulence-like density snapshot.\\
In the following analysis we choose the same parameters for our blob tracking algorithm for all runs, in order to keep comparisons between different models consistent. In order not to overestimate the velocity of individual blobs one would choose a relatively low threshold for the blob tracking algorithm. Nevertheless, the threshold cannot be set too low in this model that simulates more than one blob since it would label several independent but spatially close structures as one blob. We subtract the time and $y$-averaged radial profile from the density and apply a blob threshold of $0.2$ density units for the resulting fields. In addition, we rerun the blob tracking analysis on single isolated blobs from the previous chapter with these exact parameters to compare these two systems.

\subsection{single launch-point}
\begin{figure}[t]
	\centering
	\includegraphics[width=8cm]{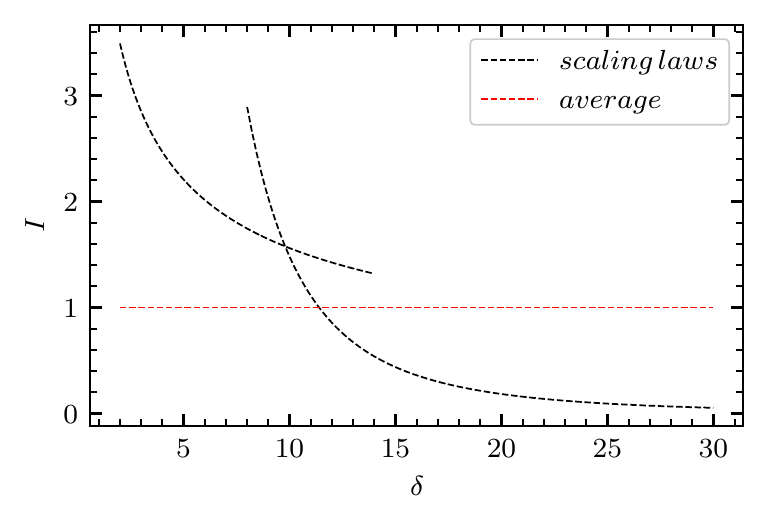}
	\caption{ Model specific intermittency parameter in dependence of blob width illustrated utilizing scaling laws for the inertial (small $\delta$) and sheath connected blob regime (big $\delta$). This is compared to the average intermittency parameter for all $\delta$. }
	\label{Intermittency}
\end{figure}

\begin{figure*}[ht]
	\centering
	\includegraphics[width=0.3\linewidth]{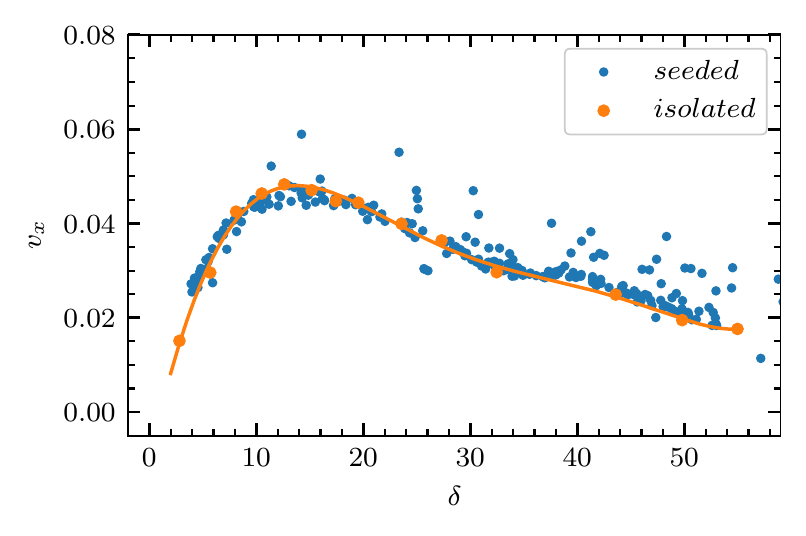}\hfil
	\includegraphics[width=0.3\linewidth]{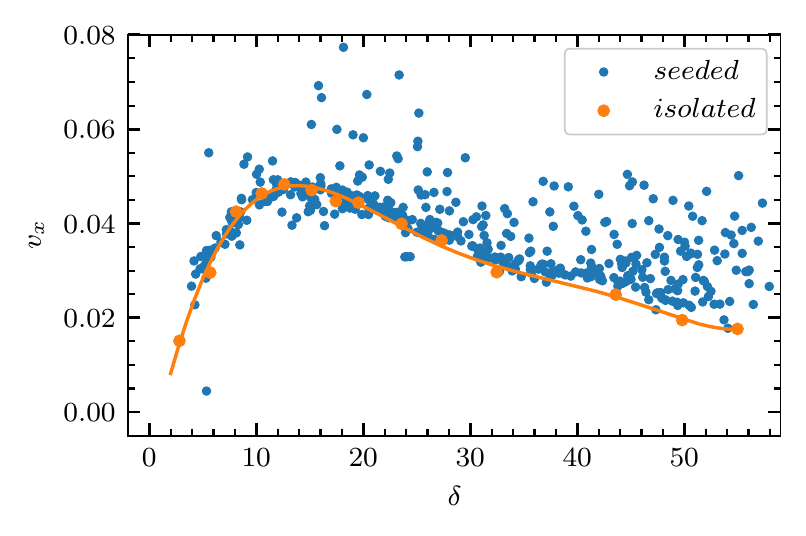}\hfil
	\includegraphics[width=0.3\linewidth]{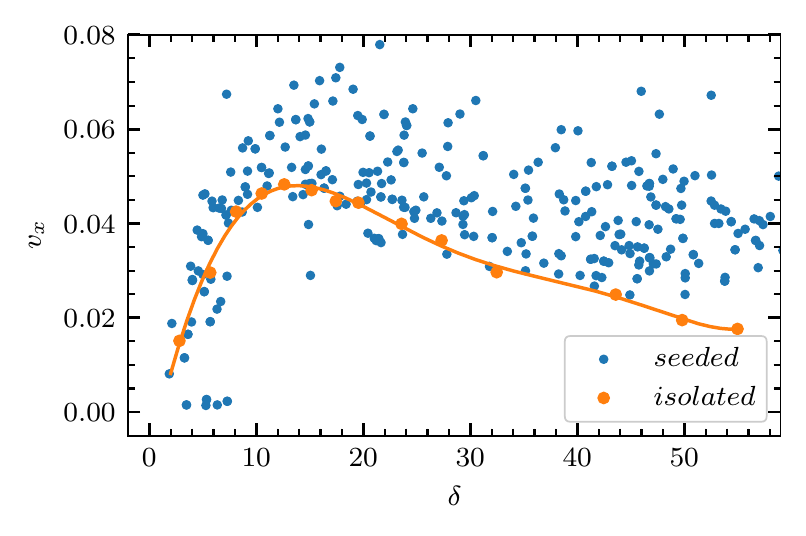}\par\medskip
	\caption {Radial velocity of randomly seeded blobs with single launch position (blue dots) compared to isolated blobs (orange dots). The intermittency parameters for the displayed runs are approximately $I = 11.8$ (left), $I = 4.9$ (middle) and $I = 1.8$ (right). Blob widths are sampled from a uniform distribution with $\delta \in U(2,30)$ and waiting times from an exponential distribution. }
	\label{single-launch}
\end{figure*}

We begin our analysis on randomly seeded blobs, keeping the blob amplitudes constant to $A=1$ and launching all blobs at $x_0 = 0.25\times L_x$ and  $y_0 = 0.5\times L_y$, which leaves the waiting times and blob widths as free parameters. In order to quantify the interaction and overlap of individual blobs we define a model specific intermittency parameter as
\begin{equation}\label{intermittency_parameter}
I = \frac{\langle v_x \rangle \langle \tau_w \rangle }{\langle \delta\rangle}
\end{equation}
where $\langle v_x \rangle$ represents the average radial velocity, $\langle \tau_w \rangle$ the average waiting time and $\langle \delta\rangle$ the average width of a specific run. This model specific intermittency parameter is introduced in the spirit of previous work on stochastic modeling of intermittent fluctuations, analyzing time series \cite{garcia2018intermittent,kube2019statistical,theodorsen2016scrape,kube2018intermittent,garcia2017sol} which defines the intermittency parameter as the ratio of the average duration time of one event above a chosen threshold, and the average waiting time between two such consecutive events. From the definition I is, strictly speaking, not constant but a function of $\delta$ of each individual blob. This effect is illustrated in figure \ref{Intermittency}, showing how the blob specific intermittency parameter deviates from the average value. This has to be taken into consideration for the following investigation. Note, that for the presented cases we calculate $\langle v_x \rangle$ and $\langle \delta\rangle$ not from input parameters of the model but from the set of seeded blobs excluding structures that only are detected for one frame.
 We launch blobs for three different average waiting times which refer to three different states of intermittency. The results of the blob tracking algorithm for these three cases are shown in figure \ref{single-launch}. 

\begin{figure*}[ht]
	\centering
	\includegraphics[width=0.4\linewidth]{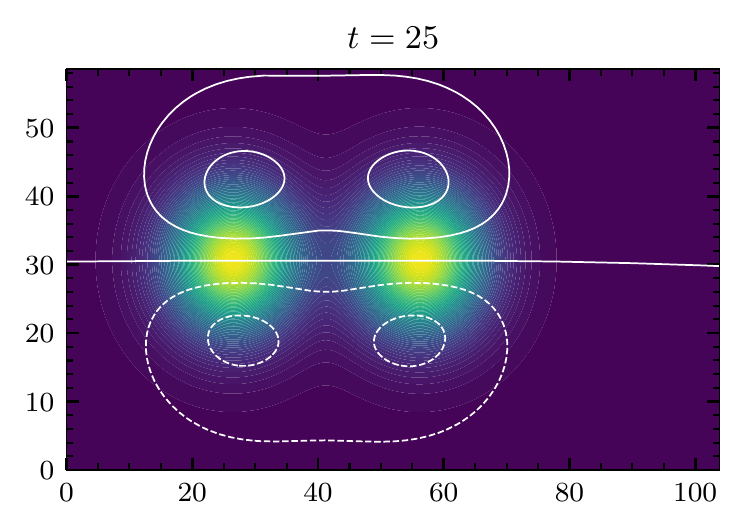}\hfil
	\includegraphics[width=0.4\linewidth]{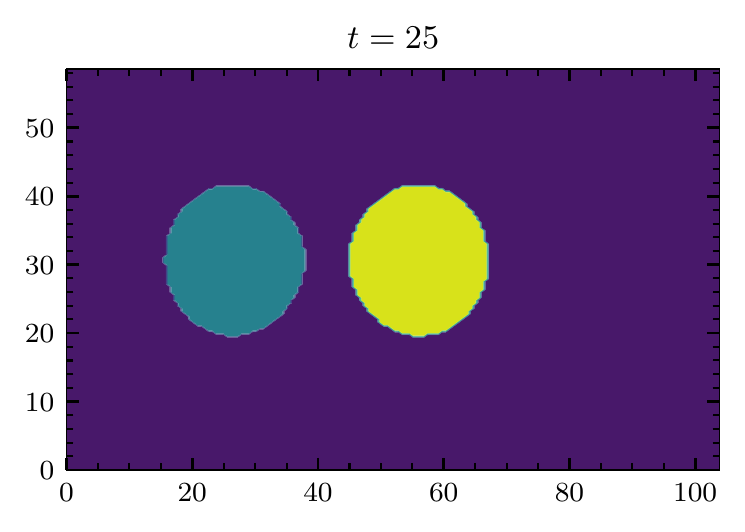}\par\medskip
	\includegraphics[width=0.4\linewidth]{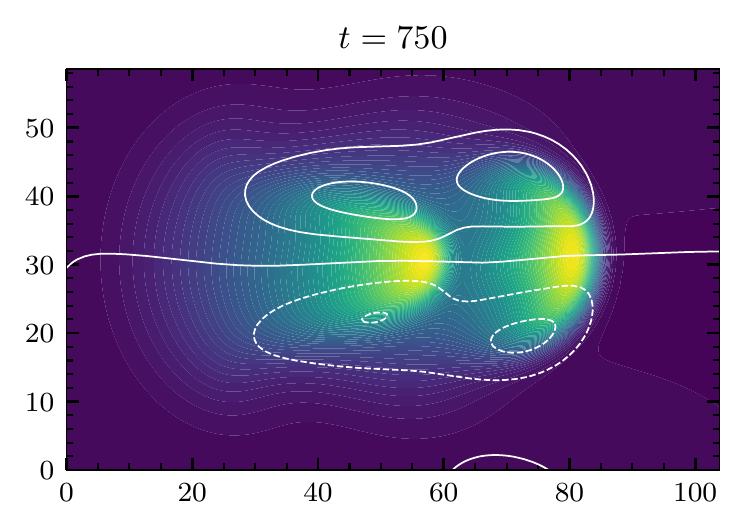}\hfil
	\includegraphics[width=0.4\linewidth]{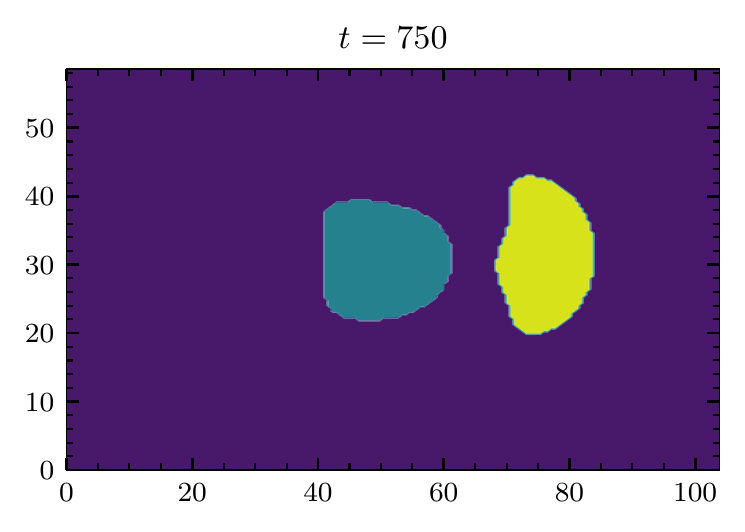}\par\medskip
	\includegraphics[width=0.4\linewidth]{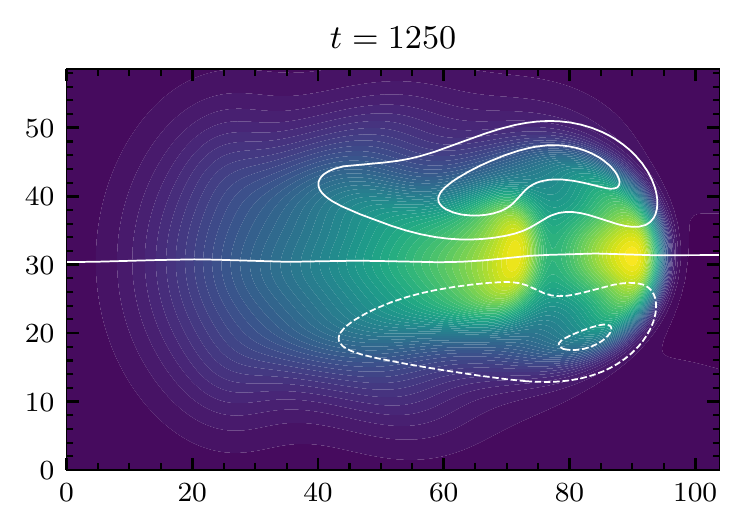}\hfil
	\includegraphics[width=0.4\linewidth]{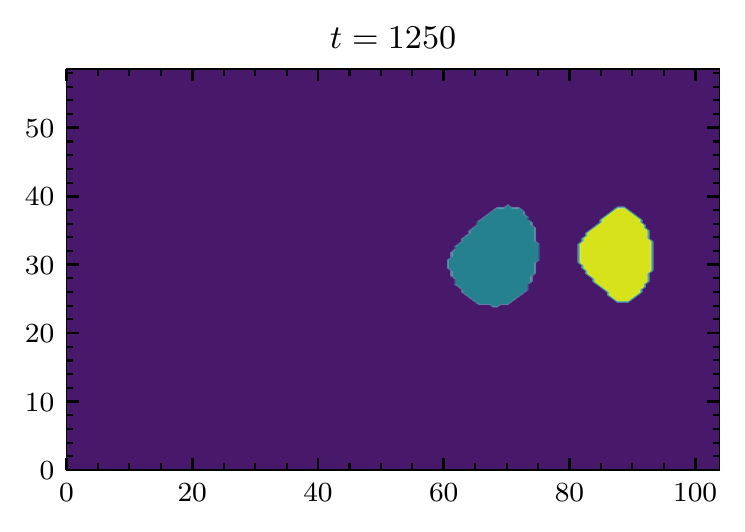}
	\caption{Snapshot of two seeded identical blobs with their electrostatic potential and the associated blob labels, detected by the blob tracking algorithm at three different time steps. The acceleration of the left blob by the electrostatic potential of the right blob is illustrated. }
	\label{two_blobs}
\end{figure*}

\begin{figure}[t]
	\centering
	\includegraphics[width=8cm]{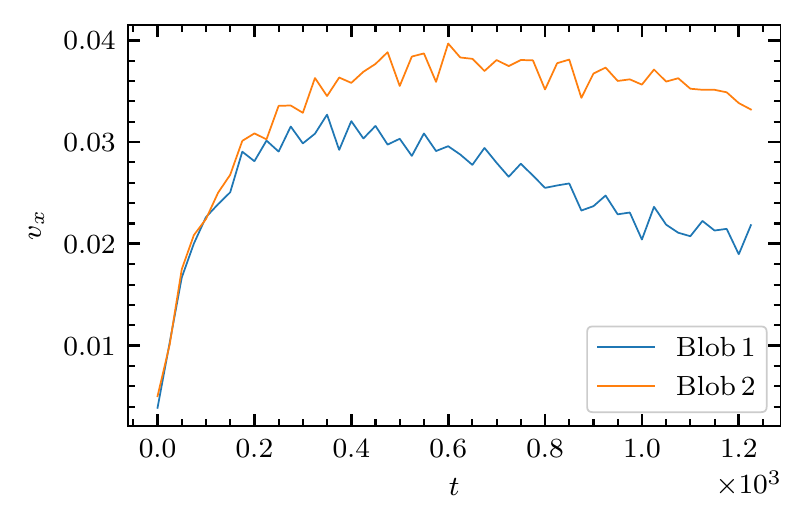}
	\caption{ Radial velocity of two seeded identical blobs at two different radial positions. Blob 2, which is trailing blob 1, shows a significant increase in the radial velocity due to the electrostatic potential created by blob 1. }
	\label{two_blobs_v}
\end{figure}

Note that the width $\delta$ of the blobs shown in figure \ref{single-launch} is determined by the blob tracking algorithm and does not exactly match the values of the input parameters. For the most intermittent case of $I = 11.8$, where blobs are the most spatially separated, we see that the overwhelming majority of detected structures lies on the line of isolated blobs. This implies that there is no strong interaction between individual blobs. Some individually detected structures show a higher radial velocity than their isolated counterparts. This effect arises due to two closely separated blobs interacting with each other's electrostatic potential. Although this has been studied in some detail in previous work \cite{militello2017interaction}, we deliver an illustration in figure \ref{two_blobs}. We seed two identical blobs at different radial positions and apply the blob tracking algorithm to determine their radial velocity. The electrostatic potential created by the two separate blobs superposes and results in a stronger electric filed which increases the \textbf{E}$\times$\textbf{B} drift that drags the coherent blob structures radially outwards. This effect leads to the formation of so called "blob trenches" in turbulence simulations. We measure radial velocity of the two blobs with the blob tracking algorithm and observe a clear increase in velocity for the second blob, shown in figure \ref{two_blobs_v}.\\

For $I = 4.9$ and $I = 1.8$ in figure \ref{single-launch}, we observe an increasing number of blobs with a higher radial velocity than their isolated counterparts. Since the average waiting time decreases, individual blobs interact strongly with the potentials of nearby blobs and get accelerated radially outwards. In addition, the blob tracking algorithm detects more smaller-sized coherent structures that usually have short lifetimes, often only one to two frames. Due to the increasing interactions and turbulent flow in this model, more of these small structures are detected by the algorithm which can be classed as numerical artifacts. For further statistical analysis these data points would usually be removed since they do not represent blobs in the conventional way. 

\subsection{random launch-point}
\begin{figure*}[ht]
	\centering
	\includegraphics[width=0.3\linewidth]{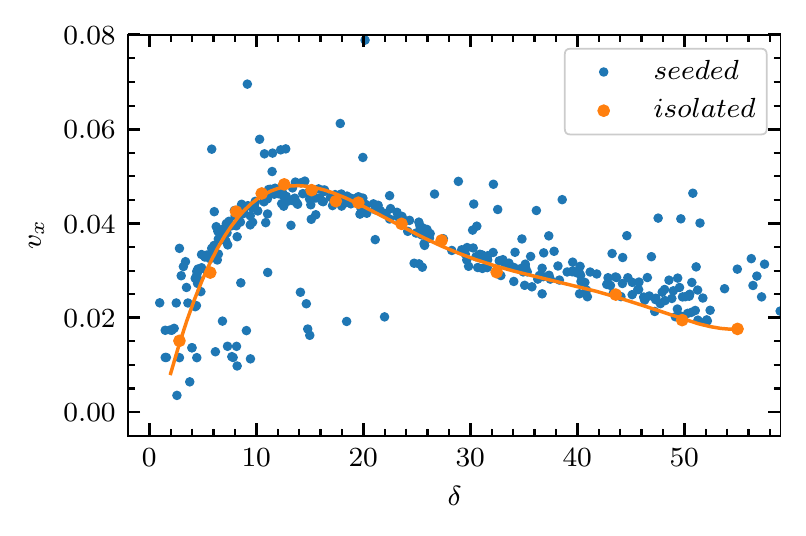}\hfil
	\includegraphics[width=0.3\linewidth]{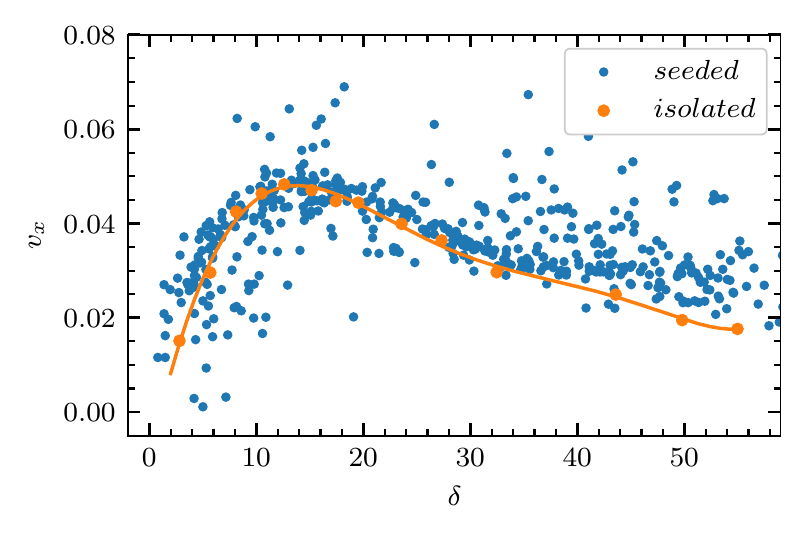}\hfil
	\includegraphics[width=0.3\linewidth]{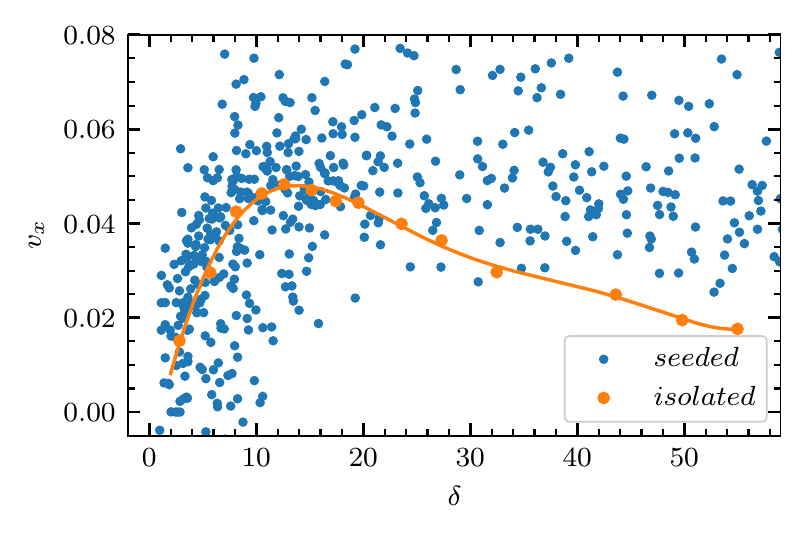}\par\medskip
	\caption {Radial velocity of randomly seeded blobs with random poloidal/binormal launch position (blue dots) are compared to isolated blobs (orange dots). The intermittency parameters for the displayed runs are approx. $I = 7.6$ (left), $I = 2.5$ (middle) and $I = 1.3$ (right).	Widths are sampled from $\delta \in U(2,30)$, initial poloidal/binormal positions from $y_0 \in U(0.2\times L_y, 0.8\times L_y)$ and waiting times from an exponential distribution. }
	\label{random_launch}
\end{figure*}

\begin{figure}[t]
	\centering
	\includegraphics[width=8cm]{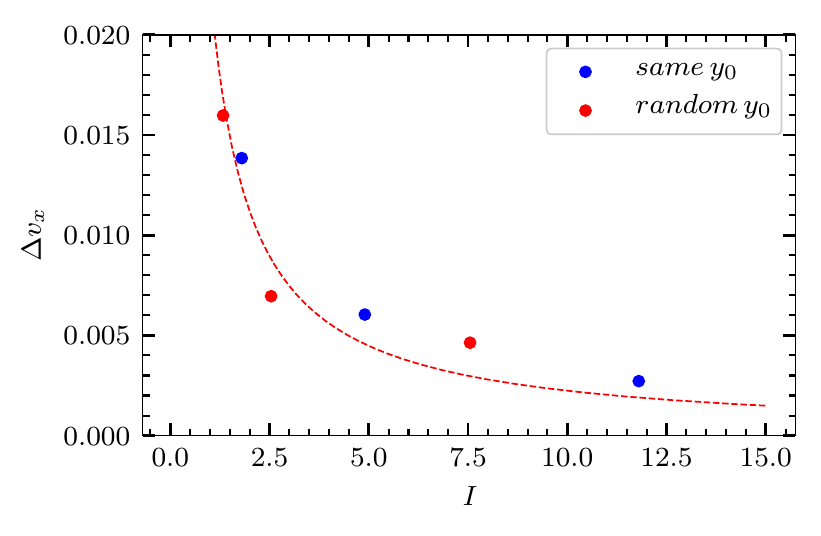}
	\caption{Average deviation in radial velocity of theoretical scaling law predictions measured in randomly seeded blob simulations for different intermittency parameters. The relationship between $\Delta v_x$ and $I$ is compared to a fit of an inverse function. }
	\label{I}
\end{figure}

The next free parameter of the investigated model added to our analysis is the poloidal/binormal launch position of the seeded blobs. We sample the launch position $y_0$ from a uniform distribution $U(0.2\times L_y, 0.8\times L_y)$ to avoid blobs propagating through the poloidal/binormal boundaries. The amplitudes remain as the last fixed parameter set to $A=1$. Seeding blobs from a random poloidal/binormal position increases the intermittency of the model and leads to more complex interactions between individual structures. We therefore multiply the expression for the intermittency parameter shown in equation \ref{intermittency_parameter} with $L_y/ 3 \langle\delta\rangle$ resulting in
\begin{equation}\label{intermittency_parameter}
I = \frac{\langle v_x \rangle \langle \tau_w \rangle L_y}{3 \langle \delta\rangle^2}
\end{equation}
to consider this extension of the model, since $L_y/3$ is the average distance of two randomly chosen events from a uniform distribution with length $L_y$. We run this model for three different intermittency parameters and present the detected blobs in figure \ref{random_launch}. 
\begin{figure}[t]
	\centering
	\includegraphics[width=8cm]{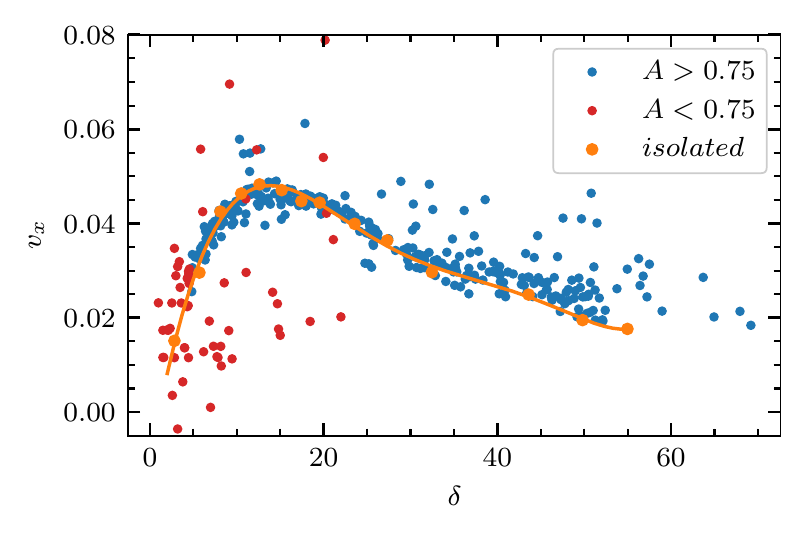}
	\caption{Radial velocity and widths of detected blobs with an intermittency parameter of $I = 7.6$ are compared to isolated blobs (orange dots). Blobs with a maximum amplitude of $A < 0.75$ (red dots) represent most small structures deviating from the scaling laws of isolated blobs. }
	\label{small_blobs}
\end{figure}
As one might expect, most detected structures in the $I = 7.6$ case follow the isolated blobs line, but show a higher spread around this line than in the single launch point model. In particular, many small blobs are detected by the blob tracking algorithm that show a significantly lower radial velocity than their isolated counterparts. We provide an explanation for this effect in figure \ref{small_blobs}. It is shown that these small blobs deviating from the theoretical predictions have a maximum amplitude significantly lower than $A=1$ as seeded blobs would have. This indicates that these small structures are not seeded by the statistical model but result from the complex interaction of seeded blobs. Since their amplitudes are significantly lower than the ones of their isolated counterparts, their radial velocity is also lower. For the cases of lower intermittency parameters we observe again an increase in the average radial velocities and the spread. This remains consistent with the previous single launch point model and can be explained by the same effects.\\\\
We utilize the presented six runs to quantify the interaction of individual blobs for different intermittency parameters. For each model we calculate the average deviation in radial velocity of the detected structures from the fit function of the isolated blobs. The result is shown in figure \ref{I}. The six data points are compared to a fit of an inverse function. This clearly suggests that the intermittency of blobs in the scrape off layer has a strong effect on their radial velocity and propagation. 

\subsection{different amplitudes}
\begin{figure*}[ht]
	\centering
	\includegraphics[width=0.3\linewidth]{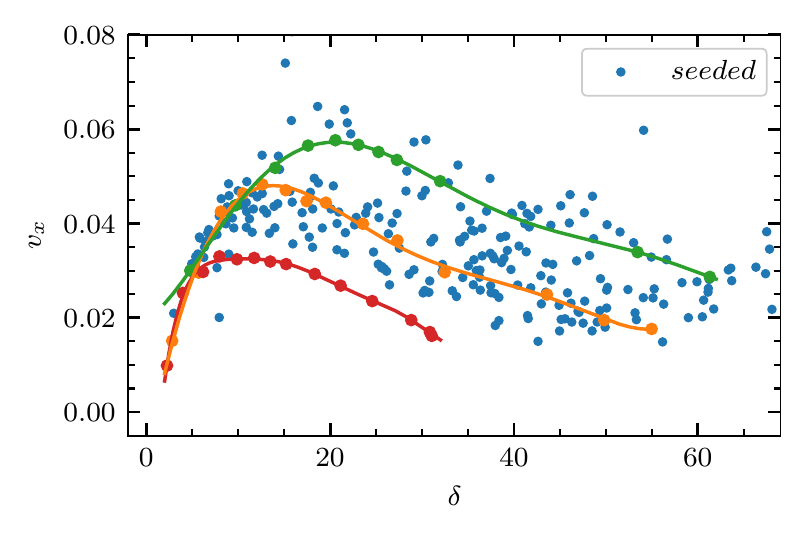}\hfil
	\includegraphics[width=0.3\linewidth]{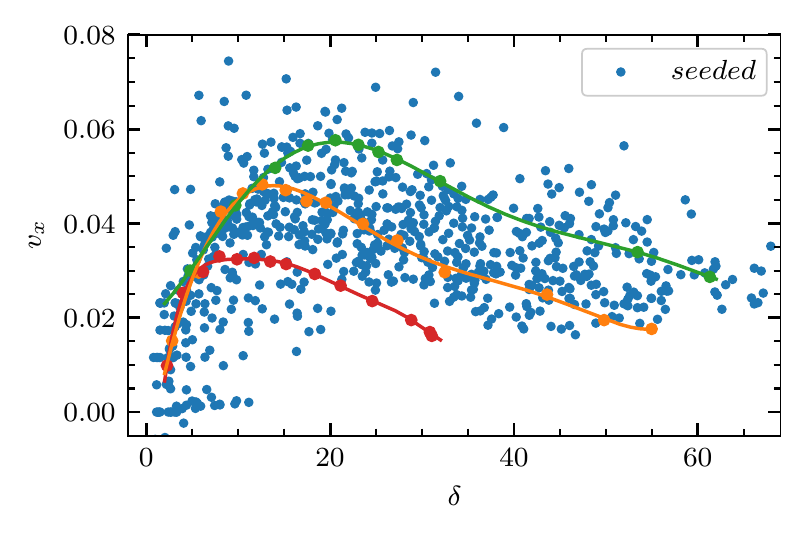}\hfil
	\includegraphics[width=0.3\linewidth]{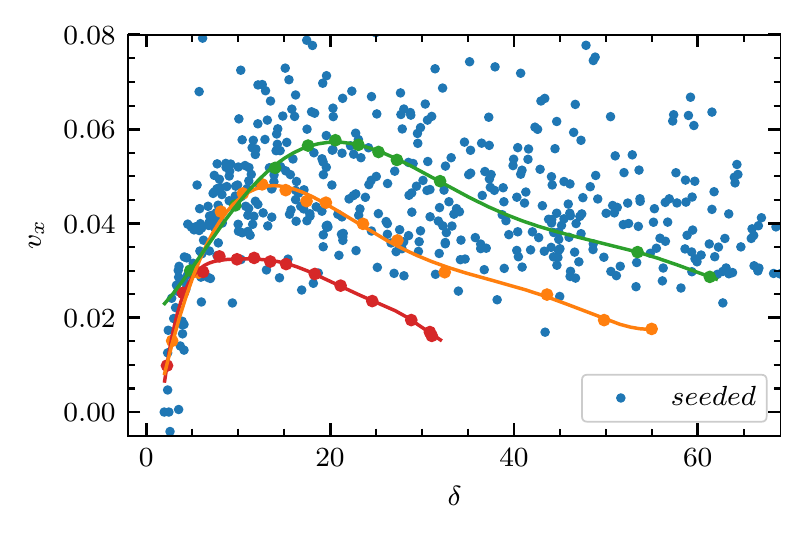}\par\medskip
	\caption{Radial velocity of randomly seeded blobs with random launch position and exponentially distributed amplitudes (blue dots) compared to isolated blobs with $A = 0.5$ (red dots), $A=1$ (orange dots) and $A=3$ (green dots). The intermittency parameters for the displayed runs are approx. $I = 10.3$ (left), $I = 7.9$ (middle) and $I = 1.7$ (right).}
	\label{different_amp}
\end{figure*}

We add the last free parameter of our model by seeding blobs with exponentially distributed amplitudes. From the sampled amplitudes we only choose those with $0.5 < A < 3$ in order to compare them more easily with isolated seeded blobs. We perform a parameter scan for blob widths for isolated seeded blobs with amplitudes $A = 0.5$ and $A=3$ in order to create reference values for the boundaries of our model. We then run our model for three different intermittency parameters and compare the results with the isolated blobs for different amplitudes. These results are shown in figure \ref{different_amp}.

\noindent The results are consistent with our previous analysis. Most randomly seeded blobs lie in between the borders established by the isolated blobs. For small blobs we observe again some data points with a lower radial velocity than in the isolated case which can be explained by the same effect as in the previous subsection. For wider structures we find some structures with higher velocities which can again be explained by the electrostatic potential of interacting blobs. As expected, the average velocity is increasing for a decreasing intermittency parameter.\\\\

\section{Turbulent simulations}
\begin{figure}[t]
	\centering
	\includegraphics[width=8cm]{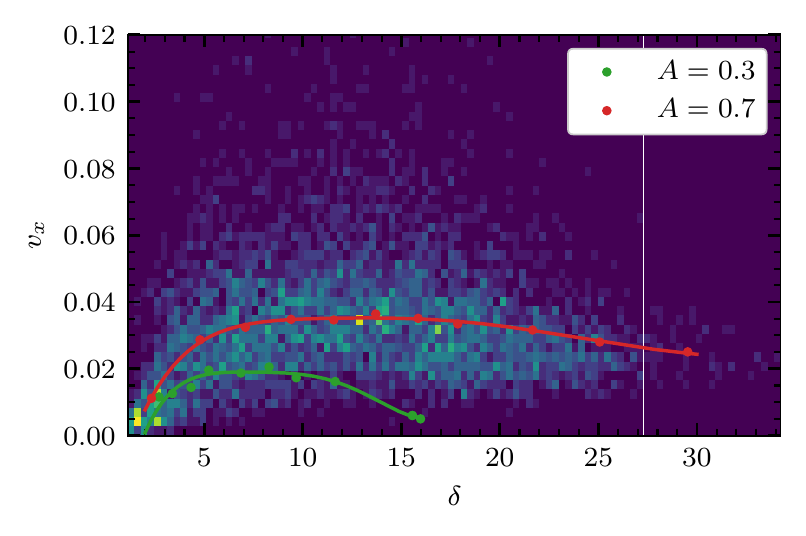}
	\caption{ Radial velocity of blobs detected in fully turbulent simulations compared to isolated blobs with $A=0.3$ and $A=0.7$. Blobs only detected for one frame are excluded as well as blobs close the poloidal/binormal simulation boundary.}
	\label{turbulence}
\end{figure}

\begin{figure*}[ht]
	\centering
	\includegraphics[width=0.3\linewidth]{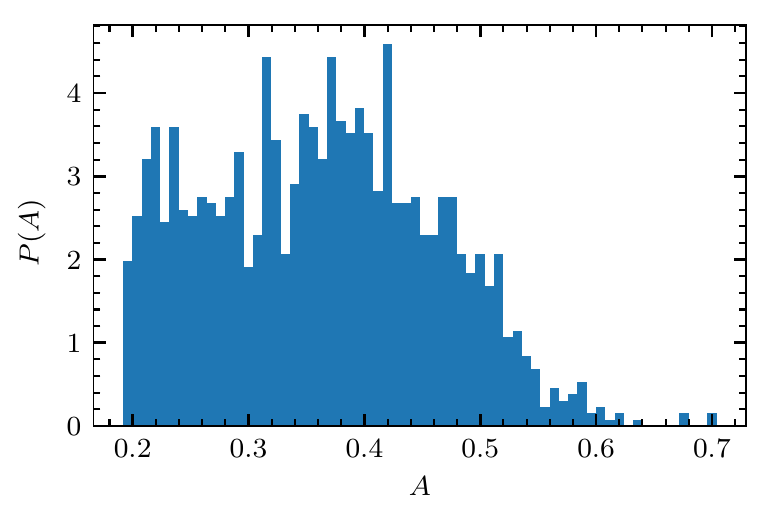}\hfil
	\includegraphics[width=0.3\linewidth]{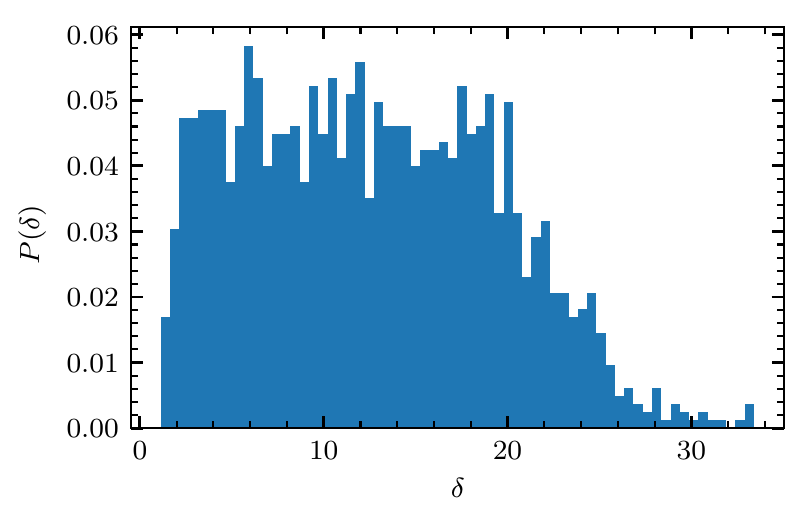}\hfil
	\includegraphics[width=0.3\linewidth]{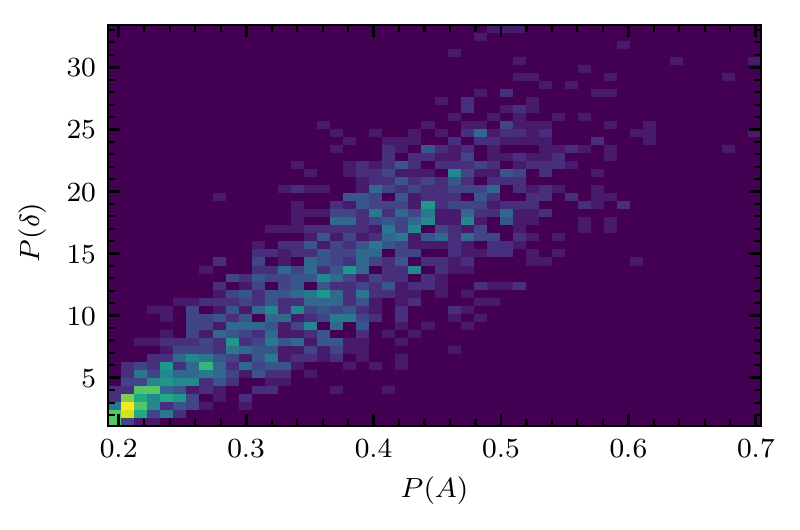}\par\medskip
	\caption{Probability density functions of maximum  amplitude (left) and  width (middle) of detected blobs in a fully turbulent simulation. The joint PDF of these parameters (right) shows the correlation between amplitude and width resulting in a correlation coefficient of $\rho = 0.85$. }
	\label{PDFs}
\end{figure*}

After investigating randomly seeded blob models, we turn our attention to a simple self consistent scrape off layer model simulating plasma turbulence. Numerically, the model stays equivalent to the seeded blob simulations but uses the term of equation \ref{source} as a plasma source instead of Gaussian seeded blobs. The density profile in the simulation domain are built and balanced by the plasma source and the sheath dissipation included in the model. These are unstable due to bad curvature and interchange instability, which leads to coherent structures of plasma propagating radially outwards due to the blob mechanism discussed in the introduction. These blob like structures vary in amplitude and width and can be detected and tracked by the tracking algorithm. \\
We exclude the source term for our blob tracking analysis and only consider coherent structures detected at $x > 0.4 \times L_x$ since this unphysical term only serves as a numerical term. In addition, we only include blobs with an initial center of mass of $0.25\times L_y < x_{init} < 0.75 \times L_y$ in our statistical evaluation in order to exclude distorted tracked structures because of the periodic boundary conditions in the $y$-dimension. Even though it is straightforward to track blobs consistently that traverse the simulation border in this direction, our numerical implementation for this issue is computationally more expensive than running the simulation longer, and only considering blobs in the central band of the domain. For such turbulence simulations the tracking algorithm identifies numerous small structures that only appear for one frame. These structures represent approximately one third of the total number of detected blobs and are also excluded in our statistical analysis. The remaining parameters for the tracking algorithm stay the same as for the randomly seeded blob model. The determined radial velocities and sizes of the detected blobs in the turbulence simulation are shown as a 2D histogram in figure \ref{turbulence}. We choose this type of plot since the illustrated 4542 blobs are too many to be shown distinctively in a scatter plot. The distribution of the sizes and amplitudes of the detected structures, as well as the joint probability distribution functions (PDF) of these two blob parameters, are shown in figure \ref{PDFs}.\\
These measurements show that the amplitudes lie in between $A=0.2$, which is equivalent to the threshold used for the blob tracking algorithm, and $A=0.7$. Since blobs with an amplitude smaller than $A=0.2$ are not detected and since many small blobs below $A=0.4$ are dissipated too quickly to be detected, the shown PDF is not representative for all structures in the system. Taking these factors into account, the common assumption of blob amplitudes being exponentially distributed cannot be falsified by these measurements even though the presented graph might indicate that. The same is valid for the distribution of blob widths. Nevertheless, we observe a clear correlation between the amplitudes and widths as the correlation coefficient of theses two parameters is $\rho = 0.85$.
In order to compare the detected blobs with their isolated counterparts we perform a parameter scan for blobs with the amplitudes $A=0.7$ and $A=0.3$ as the two edge values of the distribution. Since $A=0.2$ would be too small to be detected by the algorithm we use $A=0.3$ as the lower border. These isolated blobs are shown together with their fit in figure \ref{turbulence}. In this analysis, no blobs with a higher width than $\delta_\perp = 30$ appear, therefore we rarely observe the decreasing radial velocity for bigger and denser blobs in our velocity-size scaling. Nevertheless, the data set provides enough information to discuss the results in comparison to isolated blob simulations. As in the previous model of randomly seeded blobs with random amplitudes, we observe that the overwhelming majority of detected blob structures lie in between the trends of the isolated blob simulations. As for the previous model, the algorithm detects a significant number of structures with a higher radial velocity than the isolated blobs. We explain these events again by the interaction of blobs with the electrostatic potential of one another. Due to these findings we conclude that tracking blobs in a fully turbulent scenario shows very similar results to models of statistically seeded blobs. While the theoretical size-velocity scaling of isolated blobs gives a reasonable order of magnitude estimate, there is an order unity scatter due to strong interactions between blobs.

\section{Discussion and conclusion}

In this work we investigated the interaction of blobs in the scrape-off layer for different models of varying complexity. In particular, we compared the relation between the radial velocity and the widths of the blobs with established scaling laws. We started with studying isolated blob and extended our analysis on a model of randomly seeded blobs where the parameters are sampled from physically adequate PDFs. We studied this model for different levels of intermittency and applied the acquired knowledge on fully turbulent scrape off layer plasma simulations. \\
In this process we developed a blob tracking algorithm as a versatile tool to analyze and understand blob and plasma parameters in scrape off layer plasma simulations. We publish our implementation on github under \href{https://github.com/gregordecristoforo/xblobs}{https://github.com/gregordecristoforo/xblobs}.. The current implementation is only valid for STORM simulations but modifying the algorithm for general BOUT++, or other simulations using xarray to manage their output files, is straightforward. An extension of the algorithm to three dimensions is numerically easy to implement, but the 2D version of this algorithm can be valuable for analyzing blob propagation and turbulent transport, in a specific plane in three dimensional plasma simulations. We will use this in the future to study how blob properties depend on specific physical effects or study the plasma transport in the scrape off layer. \\
We observe an increase of the radial velocity for blobs in cases of low intermittency for the randomly seeded blob model and turbulence model, compared to isolated and intermittent cases. We explain this observation by the interaction of blobs with the electrostatic potential of one another. The blob trajectories are influenced by the electrostatic potential which gets diverted, leading to the creation of trenches in which blobs get accelerated by the potential of ones in front of them. These findings are consistent with previous work studying the interaction of two seeded blobs\cite{militello2017interaction}.  Unsurprisingly, a decrease in intermittency of the studied model shows an increase of spread in the size-velocity relation of the blobs. For all studied models we still observe a clear trend in the size-velocity relation. This concludes that despite the significant interaction of blobs, they still follow established scaling laws and can therefore be regarded to lowest order, as isolated structures propagating radially through the scrape off layer. We thereby display the relevance of isolated seeded blob and filament simulations for complex turbulent models.\\

\section*{Acknowledgements}

This work was supported with financial subvention from the Research Council of Norway under grant 240510/F20. GD acknowledges the generous hospitality of the Culham Centre for Fusion Energy (CCFE) where this work was conducted. In addition, this work has been partially funded by the EPSRC Grant EP/T012250/1 and partially carried out within the framework of the EUROfusion Consortium and has received funding from the Euratom research and training programme 2014-2018 and 2019-2020 under grant agreement No 633053. The views and opinions expressed herein do not necessarily reflect those of the European Commission. The MARCONI supercomputer was used for parts of the computational work under the project number FUA34\_SOLBOUT4.

\section*{Data Availability}
The data that support the findings of this study are available from the corresponding author upon reasonable request

\nocite{*}
\bibliography{sources}% Produces the bibliography via BibTeX.

\begin{thebibliography}{62}%
\makeatletter
\providecommand \@ifxundefined [1]{%
 \@ifx{#1\undefined}
}%
\providecommand \@ifnum [1]{%
 \ifnum #1\expandafter \@firstoftwo
 \else \expandafter \@secondoftwo
 \fi
}%
\providecommand \@ifx [1]{%
 \ifx #1\expandafter \@firstoftwo
 \else \expandafter \@secondoftwo
 \fi
}%
\providecommand \natexlab [1]{#1}%
\providecommand \enquote  [1]{``#1''}%
\providecommand \bibnamefont  [1]{#1}%
\providecommand \bibfnamefont [1]{#1}%
\providecommand \citenamefont [1]{#1}%
\providecommand \href@noop [0]{\@secondoftwo}%
\providecommand \href [0]{\begingroup \@sanitize@url \@href}%
\providecommand \@href[1]{\@@startlink{#1}\@@href}%
\providecommand \@@href[1]{\endgroup#1\@@endlink}%
\providecommand \@sanitize@url [0]{\catcode `\\12\catcode `\$12\catcode
  `\&12\catcode `\#12\catcode `\^12\catcode `\_12\catcode `\%12\relax}%
\providecommand \@@startlink[1]{}%
\providecommand \@@endlink[0]{}%
\providecommand \url  [0]{\begingroup\@sanitize@url \@url }%
\providecommand \@url [1]{\endgroup\@href {#1}{\urlprefix }}%
\providecommand \urlprefix  [0]{URL }%
\providecommand \Eprint [0]{\href }%
\providecommand \doibase [0]{http://dx.doi.org/}%
\providecommand \selectlanguage [0]{\@gobble}%
\providecommand \bibinfo  [0]{\@secondoftwo}%
\providecommand \bibfield  [0]{\@secondoftwo}%
\providecommand \translation [1]{[#1]}%
\providecommand \BibitemOpen [0]{}%
\providecommand \bibitemStop [0]{}%
\providecommand \bibitemNoStop [0]{.\EOS\space}%
\providecommand \EOS [0]{\spacefactor3000\relax}%
\providecommand \BibitemShut  [1]{\csname bibitem#1\endcsname}%
\let\auto@bib@innerbib\@empty
%</preamble>
\bibitem [{\citenamefont {Antar}\ \emph {et~al.}(2001)\citenamefont {Antar},
  \citenamefont {Krasheninnikov}, \citenamefont {Devynck}, \citenamefont
  {Doerner}, \citenamefont {Hollmann}, \citenamefont {Boedo}, \citenamefont
  {Luckhardt},\ and\ \citenamefont {Conn}}]{antar2001experimental}%
  \BibitemOpen
  \bibfield  {author} {\bibinfo {author} {\bibfnamefont {G.}~\bibnamefont
  {Antar}}, \bibinfo {author} {\bibfnamefont {S.}~\bibnamefont
  {Krasheninnikov}}, \bibinfo {author} {\bibfnamefont {P.}~\bibnamefont
  {Devynck}}, \bibinfo {author} {\bibfnamefont {R.}~\bibnamefont {Doerner}},
  \bibinfo {author} {\bibfnamefont {E.}~\bibnamefont {Hollmann}}, \bibinfo
  {author} {\bibfnamefont {J.}~\bibnamefont {Boedo}}, \bibinfo {author}
  {\bibfnamefont {S.}~\bibnamefont {Luckhardt}}, \ and\ \bibinfo {author}
  {\bibfnamefont {R.}~\bibnamefont {Conn}},\ }\bibfield  {title} {\enquote
  {\bibinfo {title} {Experimental evidence of intermittent convection in the
  edge of magnetic confinement devices},}\ }\href@noop {} {\bibfield  {journal}
  {\bibinfo  {journal} {Physical review letters}\ }\textbf {\bibinfo {volume}
  {87}},\ \bibinfo {pages} {065001} (\bibinfo {year} {2001})}\BibitemShut
  {NoStop}%
\bibitem [{\citenamefont {Antar}\ \emph {et~al.}(2003)\citenamefont {Antar},
  \citenamefont {Counsell}, \citenamefont {Yu}, \citenamefont {Labombard},\
  and\ \citenamefont {Devynck}}]{antar2003universality}%
  \BibitemOpen
  \bibfield  {author} {\bibinfo {author} {\bibfnamefont {G.~Y.}\ \bibnamefont
  {Antar}}, \bibinfo {author} {\bibfnamefont {G.}~\bibnamefont {Counsell}},
  \bibinfo {author} {\bibfnamefont {Y.}~\bibnamefont {Yu}}, \bibinfo {author}
  {\bibfnamefont {B.}~\bibnamefont {Labombard}}, \ and\ \bibinfo {author}
  {\bibfnamefont {P.}~\bibnamefont {Devynck}},\ }\bibfield  {title} {\enquote
  {\bibinfo {title} {Universality of intermittent convective transport in the
  scrape-off layer of magnetically confined devices},}\ }\href@noop {}
  {\bibfield  {journal} {\bibinfo  {journal} {Physics of Plasmas}\ }\textbf
  {\bibinfo {volume} {10}},\ \bibinfo {pages} {419--428} (\bibinfo {year}
  {2003})}\BibitemShut {NoStop}%
\bibitem [{\citenamefont {Kirk}\ \emph {et~al.}(2006)\citenamefont {Kirk},
  \citenamefont {Ayed}, \citenamefont {Counsell}, \citenamefont {Dudson},
  \citenamefont {Eich}, \citenamefont {Herrmann}, \citenamefont {Koch},
  \citenamefont {Martin}, \citenamefont {Meakins}, \citenamefont {Saarelma}
  \emph {et~al.}}]{kirk2006filament}%
  \BibitemOpen
  \bibfield  {author} {\bibinfo {author} {\bibfnamefont {A.}~\bibnamefont
  {Kirk}}, \bibinfo {author} {\bibfnamefont {N.~B.}\ \bibnamefont {Ayed}},
  \bibinfo {author} {\bibfnamefont {G.}~\bibnamefont {Counsell}}, \bibinfo
  {author} {\bibfnamefont {B.}~\bibnamefont {Dudson}}, \bibinfo {author}
  {\bibfnamefont {T.}~\bibnamefont {Eich}}, \bibinfo {author} {\bibfnamefont
  {A.}~\bibnamefont {Herrmann}}, \bibinfo {author} {\bibfnamefont
  {B.}~\bibnamefont {Koch}}, \bibinfo {author} {\bibfnamefont {R.}~\bibnamefont
  {Martin}}, \bibinfo {author} {\bibfnamefont {A.}~\bibnamefont {Meakins}},
  \bibinfo {author} {\bibfnamefont {S.}~\bibnamefont {Saarelma}},  \emph
  {et~al.},\ }\bibfield  {title} {\enquote {\bibinfo {title} {Filament
  structures at the plasma edge on mast},}\ }\href@noop {} {\bibfield
  {journal} {\bibinfo  {journal} {Plasma physics and controlled fusion}\
  }\textbf {\bibinfo {volume} {48}},\ \bibinfo {pages} {B433} (\bibinfo {year}
  {2006})}\BibitemShut {NoStop}%
\bibitem [{\citenamefont {Dudson}\ \emph {et~al.}(2008)\citenamefont {Dudson},
  \citenamefont {Ayed}, \citenamefont {Kirk}, \citenamefont {Wilson},
  \citenamefont {Counsell}, \citenamefont {Xu}, \citenamefont {Umansky},
  \citenamefont {Snyder}, \citenamefont {LLoyd} \emph
  {et~al.}}]{dudson2008experiments}%
  \BibitemOpen
  \bibfield  {author} {\bibinfo {author} {\bibfnamefont {B.}~\bibnamefont
  {Dudson}}, \bibinfo {author} {\bibfnamefont {N.~B.}\ \bibnamefont {Ayed}},
  \bibinfo {author} {\bibfnamefont {A.}~\bibnamefont {Kirk}}, \bibinfo {author}
  {\bibfnamefont {H.}~\bibnamefont {Wilson}}, \bibinfo {author} {\bibfnamefont
  {G.}~\bibnamefont {Counsell}}, \bibinfo {author} {\bibfnamefont
  {X.}~\bibnamefont {Xu}}, \bibinfo {author} {\bibfnamefont {M.}~\bibnamefont
  {Umansky}}, \bibinfo {author} {\bibfnamefont {P.}~\bibnamefont {Snyder}},
  \bibinfo {author} {\bibfnamefont {B.}~\bibnamefont {LLoyd}},  \emph
  {et~al.},\ }\bibfield  {title} {\enquote {\bibinfo {title} {Experiments and
  simulation of edge turbulence and filaments in mast},}\ }\href@noop {}
  {\bibfield  {journal} {\bibinfo  {journal} {Plasma Physics and Controlled
  Fusion}\ }\textbf {\bibinfo {volume} {50}},\ \bibinfo {pages} {124012}
  (\bibinfo {year} {2008})}\BibitemShut {NoStop}%
\bibitem [{\citenamefont {Ayed}\ \emph {et~al.}(2009)\citenamefont {Ayed},
  \citenamefont {Kirk}, \citenamefont {Dudson}, \citenamefont {Tallents},
  \citenamefont {Vann}, \citenamefont {Wilson} \emph {et~al.}}]{ayed2009inter}%
  \BibitemOpen
  \bibfield  {author} {\bibinfo {author} {\bibfnamefont {N.~B.}\ \bibnamefont
  {Ayed}}, \bibinfo {author} {\bibfnamefont {A.}~\bibnamefont {Kirk}}, \bibinfo
  {author} {\bibfnamefont {B.}~\bibnamefont {Dudson}}, \bibinfo {author}
  {\bibfnamefont {S.}~\bibnamefont {Tallents}}, \bibinfo {author}
  {\bibfnamefont {R.}~\bibnamefont {Vann}}, \bibinfo {author} {\bibfnamefont
  {H.}~\bibnamefont {Wilson}},  \emph {et~al.},\ }\bibfield  {title} {\enquote
  {\bibinfo {title} {Inter-elm filaments and turbulent transport in the
  mega-amp spherical tokamak},}\ }\href@noop {} {\bibfield  {journal} {\bibinfo
   {journal} {Plasma Physics and Controlled Fusion}\ }\textbf {\bibinfo
  {volume} {51}},\ \bibinfo {pages} {035016} (\bibinfo {year}
  {2009})}\BibitemShut {NoStop}%
\bibitem [{\citenamefont {Boedo}\ \emph {et~al.}(2001)\citenamefont {Boedo},
  \citenamefont {Rudakov}, \citenamefont {Moyer}, \citenamefont
  {Krasheninnikov}, \citenamefont {Whyte}, \citenamefont {McKee}, \citenamefont
  {Tynan}, \citenamefont {Schaffer}, \citenamefont {Stangeby}, \citenamefont
  {West} \emph {et~al.}}]{boedo2001transport}%
  \BibitemOpen
  \bibfield  {author} {\bibinfo {author} {\bibfnamefont {J.~A.}\ \bibnamefont
  {Boedo}}, \bibinfo {author} {\bibfnamefont {D.}~\bibnamefont {Rudakov}},
  \bibinfo {author} {\bibfnamefont {R.}~\bibnamefont {Moyer}}, \bibinfo
  {author} {\bibfnamefont {S.}~\bibnamefont {Krasheninnikov}}, \bibinfo
  {author} {\bibfnamefont {D.}~\bibnamefont {Whyte}}, \bibinfo {author}
  {\bibfnamefont {G.}~\bibnamefont {McKee}}, \bibinfo {author} {\bibfnamefont
  {G.}~\bibnamefont {Tynan}}, \bibinfo {author} {\bibfnamefont
  {M.}~\bibnamefont {Schaffer}}, \bibinfo {author} {\bibfnamefont
  {P.}~\bibnamefont {Stangeby}}, \bibinfo {author} {\bibfnamefont
  {P.}~\bibnamefont {West}},  \emph {et~al.},\ }\bibfield  {title} {\enquote
  {\bibinfo {title} {Transport by intermittent convection in the boundary of
  the diii-d tokamak},}\ }\href@noop {} {\bibfield  {journal} {\bibinfo
  {journal} {Physics of Plasmas}\ }\textbf {\bibinfo {volume} {8}},\ \bibinfo
  {pages} {4826--4833} (\bibinfo {year} {2001})}\BibitemShut {NoStop}%
\bibitem [{\citenamefont {Rudakov}\ \emph {et~al.}(2002)\citenamefont
  {Rudakov}, \citenamefont {Boedo}, \citenamefont {Moyer}, \citenamefont
  {Krasheninnikov}, \citenamefont {Leonard}, \citenamefont {Mahdavi},
  \citenamefont {McKee}, \citenamefont {Porter}, \citenamefont {Stangeby},
  \citenamefont {Watkins} \emph {et~al.}}]{rudakov2002fluctuation}%
  \BibitemOpen
  \bibfield  {author} {\bibinfo {author} {\bibfnamefont {D.}~\bibnamefont
  {Rudakov}}, \bibinfo {author} {\bibfnamefont {J.}~\bibnamefont {Boedo}},
  \bibinfo {author} {\bibfnamefont {R.}~\bibnamefont {Moyer}}, \bibinfo
  {author} {\bibfnamefont {S.}~\bibnamefont {Krasheninnikov}}, \bibinfo
  {author} {\bibfnamefont {A.}~\bibnamefont {Leonard}}, \bibinfo {author}
  {\bibfnamefont {M.}~\bibnamefont {Mahdavi}}, \bibinfo {author} {\bibfnamefont
  {G.}~\bibnamefont {McKee}}, \bibinfo {author} {\bibfnamefont
  {G.}~\bibnamefont {Porter}}, \bibinfo {author} {\bibfnamefont
  {P.}~\bibnamefont {Stangeby}}, \bibinfo {author} {\bibfnamefont
  {J.}~\bibnamefont {Watkins}},  \emph {et~al.},\ }\bibfield  {title} {\enquote
  {\bibinfo {title} {Fluctuation-driven transport in the diii-d boundary},}\
  }\href@noop {} {\bibfield  {journal} {\bibinfo  {journal} {Plasma physics and
  controlled fusion}\ }\textbf {\bibinfo {volume} {44}},\ \bibinfo {pages}
  {717} (\bibinfo {year} {2002})}\BibitemShut {NoStop}%
\bibitem [{\citenamefont {Boedo}\ \emph {et~al.}(2003)\citenamefont {Boedo},
  \citenamefont {Rudakov}, \citenamefont {Moyer}, \citenamefont {McKee},
  \citenamefont {Colchin}, \citenamefont {Schaffer}, \citenamefont {Stangeby},
  \citenamefont {West}, \citenamefont {Allen}, \citenamefont {Evans} \emph
  {et~al.}}]{boedo2003transport}%
  \BibitemOpen
  \bibfield  {author} {\bibinfo {author} {\bibfnamefont {J.~A.}\ \bibnamefont
  {Boedo}}, \bibinfo {author} {\bibfnamefont {D.~L.}\ \bibnamefont {Rudakov}},
  \bibinfo {author} {\bibfnamefont {R.~A.}\ \bibnamefont {Moyer}}, \bibinfo
  {author} {\bibfnamefont {G.~R.}\ \bibnamefont {McKee}}, \bibinfo {author}
  {\bibfnamefont {R.~J.}\ \bibnamefont {Colchin}}, \bibinfo {author}
  {\bibfnamefont {M.~J.}\ \bibnamefont {Schaffer}}, \bibinfo {author}
  {\bibfnamefont {P.}~\bibnamefont {Stangeby}}, \bibinfo {author}
  {\bibfnamefont {W.}~\bibnamefont {West}}, \bibinfo {author} {\bibfnamefont
  {S.~L.}\ \bibnamefont {Allen}}, \bibinfo {author} {\bibfnamefont {T.~E.}\
  \bibnamefont {Evans}},  \emph {et~al.},\ }\bibfield  {title} {\enquote
  {\bibinfo {title} {Transport by intermittency in the boundary of the diii-d
  tokamak},}\ }\href@noop {} {\bibfield  {journal} {\bibinfo  {journal}
  {Physics of Plasmas}\ }\textbf {\bibinfo {volume} {10}},\ \bibinfo {pages}
  {1670--1677} (\bibinfo {year} {2003})}\BibitemShut {NoStop}%
\bibitem [{\citenamefont {Garcia}\ \emph
  {et~al.}(2007{\natexlab{a}})\citenamefont {Garcia}, \citenamefont {Horacek},
  \citenamefont {Pitts}, \citenamefont {Nielsen}, \citenamefont {Fundamenski},
  \citenamefont {Naulin},\ and\ \citenamefont
  {Rasmussen}}]{garcia2007fluctuations}%
  \BibitemOpen
  \bibfield  {author} {\bibinfo {author} {\bibfnamefont {O.~E.}\ \bibnamefont
  {Garcia}}, \bibinfo {author} {\bibfnamefont {J.}~\bibnamefont {Horacek}},
  \bibinfo {author} {\bibfnamefont {R.}~\bibnamefont {Pitts}}, \bibinfo
  {author} {\bibfnamefont {A.~H.}\ \bibnamefont {Nielsen}}, \bibinfo {author}
  {\bibfnamefont {W.}~\bibnamefont {Fundamenski}}, \bibinfo {author}
  {\bibfnamefont {V.}~\bibnamefont {Naulin}}, \ and\ \bibinfo {author}
  {\bibfnamefont {J.~J.}\ \bibnamefont {Rasmussen}},\ }\bibfield  {title}
  {\enquote {\bibinfo {title} {Fluctuations and transport in the tcv scrape-off
  layer},}\ }\href@noop {} {\bibfield  {journal} {\bibinfo  {journal} {Nuclear
  fusion}\ }\textbf {\bibinfo {volume} {47}},\ \bibinfo {pages} {667} (\bibinfo
  {year} {2007}{\natexlab{a}})}\BibitemShut {NoStop}%
\bibitem [{\citenamefont {Garcia}\ \emph
  {et~al.}(2007{\natexlab{b}})\citenamefont {Garcia}, \citenamefont {Pitts},
  \citenamefont {Horacek}, \citenamefont {Madsen}, \citenamefont {Naulin},
  \citenamefont {Nielsen},\ and\ \citenamefont
  {Rasmussen}}]{garcia2007collisionality}%
  \BibitemOpen
  \bibfield  {author} {\bibinfo {author} {\bibfnamefont {O.~E.}\ \bibnamefont
  {Garcia}}, \bibinfo {author} {\bibfnamefont {R.}~\bibnamefont {Pitts}},
  \bibinfo {author} {\bibfnamefont {J.}~\bibnamefont {Horacek}}, \bibinfo
  {author} {\bibfnamefont {J.}~\bibnamefont {Madsen}}, \bibinfo {author}
  {\bibfnamefont {V.}~\bibnamefont {Naulin}}, \bibinfo {author} {\bibfnamefont
  {A.~H.}\ \bibnamefont {Nielsen}}, \ and\ \bibinfo {author} {\bibfnamefont
  {J.~J.}\ \bibnamefont {Rasmussen}},\ }\bibfield  {title} {\enquote {\bibinfo
  {title} {Collisionality dependent transport in tcv sol plasmas},}\
  }\href@noop {} {\bibfield  {journal} {\bibinfo  {journal} {Plasma Physics and
  Controlled Fusion}\ }\textbf {\bibinfo {volume} {49}},\ \bibinfo {pages}
  {B47} (\bibinfo {year} {2007}{\natexlab{b}})}\BibitemShut {NoStop}%
\bibitem [{\citenamefont {Militello}\ \emph {et~al.}(2013)\citenamefont
  {Militello}, \citenamefont {Tamain}, \citenamefont {Fundamenski},
  \citenamefont {Kirk}, \citenamefont {Naulin}, \citenamefont {Nielsen} \emph
  {et~al.}}]{militello2013experimental}%
  \BibitemOpen
  \bibfield  {author} {\bibinfo {author} {\bibfnamefont {F.}~\bibnamefont
  {Militello}}, \bibinfo {author} {\bibfnamefont {P.}~\bibnamefont {Tamain}},
  \bibinfo {author} {\bibfnamefont {W.}~\bibnamefont {Fundamenski}}, \bibinfo
  {author} {\bibfnamefont {A.}~\bibnamefont {Kirk}}, \bibinfo {author}
  {\bibfnamefont {V.}~\bibnamefont {Naulin}}, \bibinfo {author} {\bibfnamefont
  {A.~H.}\ \bibnamefont {Nielsen}},  \emph {et~al.},\ }\bibfield  {title}
  {\enquote {\bibinfo {title} {Experimental and numerical characterization of
  the turbulence in the scrape-off layer of mast},}\ }\href@noop {} {\bibfield
  {journal} {\bibinfo  {journal} {Plasma Physics and Controlled Fusion}\
  }\textbf {\bibinfo {volume} {55}},\ \bibinfo {pages} {025005} (\bibinfo
  {year} {2013})}\BibitemShut {NoStop}%
\bibitem [{\citenamefont {Farley}\ \emph {et~al.}(2017)\citenamefont {Farley},
  \citenamefont {Militello}, \citenamefont {Walkden}, \citenamefont {Harrison},
  \citenamefont {Silburn},\ and\ \citenamefont {Bradley}}]{farley2017analysis}%
  \BibitemOpen
  \bibfield  {author} {\bibinfo {author} {\bibfnamefont {T.}~\bibnamefont
  {Farley}}, \bibinfo {author} {\bibfnamefont {F.}~\bibnamefont {Militello}},
  \bibinfo {author} {\bibfnamefont {N.}~\bibnamefont {Walkden}}, \bibinfo
  {author} {\bibfnamefont {J.}~\bibnamefont {Harrison}}, \bibinfo {author}
  {\bibfnamefont {S.}~\bibnamefont {Silburn}}, \ and\ \bibinfo {author}
  {\bibfnamefont {J.}~\bibnamefont {Bradley}},\ }\bibfield  {title} {\enquote
  {\bibinfo {title} {Analysis of filament statistics in fast camera data on
  mast},}\ }in\ \href@noop {} {\emph {\bibinfo {booktitle} {APS Meeting
  Abstracts}}}\ (\bibinfo {year} {2017})\BibitemShut {NoStop}%
\bibitem [{\citenamefont {Walkden}\ \emph
  {et~al.}(2017{\natexlab{a}})\citenamefont {Walkden}, \citenamefont
  {Harrison}, \citenamefont {Silburn}, \citenamefont {Farley}, \citenamefont
  {Henderson}, \citenamefont {Kirk}, \citenamefont {Militello}, \citenamefont
  {Thornton}, \citenamefont {Team} \emph {et~al.}}]{walkden2017quiescence}%
  \BibitemOpen
  \bibfield  {author} {\bibinfo {author} {\bibfnamefont {N.}~\bibnamefont
  {Walkden}}, \bibinfo {author} {\bibfnamefont {J.}~\bibnamefont {Harrison}},
  \bibinfo {author} {\bibfnamefont {S.}~\bibnamefont {Silburn}}, \bibinfo
  {author} {\bibfnamefont {T.}~\bibnamefont {Farley}}, \bibinfo {author}
  {\bibfnamefont {S.~S.}\ \bibnamefont {Henderson}}, \bibinfo {author}
  {\bibfnamefont {A.}~\bibnamefont {Kirk}}, \bibinfo {author} {\bibfnamefont
  {F.}~\bibnamefont {Militello}}, \bibinfo {author} {\bibfnamefont
  {A.}~\bibnamefont {Thornton}}, \bibinfo {author} {\bibfnamefont
  {M.}~\bibnamefont {Team}},  \emph {et~al.},\ }\bibfield  {title} {\enquote
  {\bibinfo {title} {Quiescence near the x-point of mast measured by high speed
  visible imaging},}\ }\href@noop {} {\bibfield  {journal} {\bibinfo  {journal}
  {Nuclear Fusion}\ }\textbf {\bibinfo {volume} {57}},\ \bibinfo {pages}
  {126028} (\bibinfo {year} {2017}{\natexlab{a}})}\BibitemShut {NoStop}%
\bibitem [{\citenamefont {Walkden}\ \emph
  {et~al.}(2017{\natexlab{b}})\citenamefont {Walkden}, \citenamefont
  {Militello}, \citenamefont {Harrison}, \citenamefont {Farley}, \citenamefont
  {Silburn},\ and\ \citenamefont {Young}}]{walkden2017identification}%
  \BibitemOpen
  \bibfield  {author} {\bibinfo {author} {\bibfnamefont {N.}~\bibnamefont
  {Walkden}}, \bibinfo {author} {\bibfnamefont {F.}~\bibnamefont {Militello}},
  \bibinfo {author} {\bibfnamefont {J.}~\bibnamefont {Harrison}}, \bibinfo
  {author} {\bibfnamefont {T.}~\bibnamefont {Farley}}, \bibinfo {author}
  {\bibfnamefont {S.}~\bibnamefont {Silburn}}, \ and\ \bibinfo {author}
  {\bibfnamefont {J.}~\bibnamefont {Young}},\ }\bibfield  {title} {\enquote
  {\bibinfo {title} {Identification of intermittent transport in the scrape-off
  layer of mast through high speed imaging},}\ }\href@noop {} {\bibfield
  {journal} {\bibinfo  {journal} {Nuclear Materials and Energy}\ }\textbf
  {\bibinfo {volume} {12}},\ \bibinfo {pages} {175--180} (\bibinfo {year}
  {2017}{\natexlab{b}})}\BibitemShut {NoStop}%
\bibitem [{\citenamefont {Walkden}\ \emph {et~al.}(2018)\citenamefont
  {Walkden}, \citenamefont {Labit}, \citenamefont {Reimerdes}, \citenamefont
  {Harrison}, \citenamefont {Farley}, \citenamefont {Innocente}, \citenamefont
  {Militello}, \citenamefont {Team}, \citenamefont {Team} \emph
  {et~al.}}]{walkden2018fluctuation}%
  \BibitemOpen
  \bibfield  {author} {\bibinfo {author} {\bibfnamefont {N.}~\bibnamefont
  {Walkden}}, \bibinfo {author} {\bibfnamefont {B.}~\bibnamefont {Labit}},
  \bibinfo {author} {\bibfnamefont {H.}~\bibnamefont {Reimerdes}}, \bibinfo
  {author} {\bibfnamefont {J.}~\bibnamefont {Harrison}}, \bibinfo {author}
  {\bibfnamefont {T.}~\bibnamefont {Farley}}, \bibinfo {author} {\bibfnamefont
  {P.}~\bibnamefont {Innocente}}, \bibinfo {author} {\bibfnamefont
  {F.}~\bibnamefont {Militello}}, \bibinfo {author} {\bibfnamefont
  {T.}~\bibnamefont {Team}}, \bibinfo {author} {\bibfnamefont {M.}~\bibnamefont
  {Team}},  \emph {et~al.},\ }\bibfield  {title} {\enquote {\bibinfo {title}
  {Fluctuation characteristics of the tcv snowflake divertor measured with high
  speed visible imaging},}\ }\href@noop {} {\bibfield  {journal} {\bibinfo
  {journal} {Plasma Physics and Controlled Fusion}\ }\textbf {\bibinfo {volume}
  {60}},\ \bibinfo {pages} {115008} (\bibinfo {year} {2018})}\BibitemShut
  {NoStop}%
\bibitem [{\citenamefont {Zweben}\ \emph {et~al.}(2002)\citenamefont {Zweben},
  \citenamefont {Stotler}, \citenamefont {Terry}, \citenamefont {LaBombard},
  \citenamefont {Greenwald}, \citenamefont {Muterspaugh}, \citenamefont
  {Pitcher}, \citenamefont {Group}, \citenamefont {Hallatschek}, \citenamefont
  {Maqueda} \emph {et~al.}}]{zweben2002edge}%
  \BibitemOpen
  \bibfield  {author} {\bibinfo {author} {\bibfnamefont {S.}~\bibnamefont
  {Zweben}}, \bibinfo {author} {\bibfnamefont {D.}~\bibnamefont {Stotler}},
  \bibinfo {author} {\bibfnamefont {J.}~\bibnamefont {Terry}}, \bibinfo
  {author} {\bibfnamefont {B.}~\bibnamefont {LaBombard}}, \bibinfo {author}
  {\bibfnamefont {M.}~\bibnamefont {Greenwald}}, \bibinfo {author}
  {\bibfnamefont {M.}~\bibnamefont {Muterspaugh}}, \bibinfo {author}
  {\bibfnamefont {C.}~\bibnamefont {Pitcher}}, \bibinfo {author} {\bibfnamefont
  {A.~C.-M.}\ \bibnamefont {Group}}, \bibinfo {author} {\bibfnamefont
  {K.}~\bibnamefont {Hallatschek}}, \bibinfo {author} {\bibfnamefont
  {R.}~\bibnamefont {Maqueda}},  \emph {et~al.},\ }\bibfield  {title} {\enquote
  {\bibinfo {title} {Edge turbulence imaging in the alcator c-mod tokamak},}\
  }\href@noop {} {\bibfield  {journal} {\bibinfo  {journal} {Physics of
  Plasmas}\ }\textbf {\bibinfo {volume} {9}},\ \bibinfo {pages} {1981--1989}
  (\bibinfo {year} {2002})}\BibitemShut {NoStop}%
\bibitem [{\citenamefont {Terry}\ \emph {et~al.}(2003)\citenamefont {Terry},
  \citenamefont {Zweben}, \citenamefont {Hallatschek}, \citenamefont
  {LaBombard}, \citenamefont {Maqueda}, \citenamefont {Bai}, \citenamefont
  {Boswell}, \citenamefont {Greenwald}, \citenamefont {Kopon}, \citenamefont
  {Nevins} \emph {et~al.}}]{terry2003observations}%
  \BibitemOpen
  \bibfield  {author} {\bibinfo {author} {\bibfnamefont {J.}~\bibnamefont
  {Terry}}, \bibinfo {author} {\bibfnamefont {S.}~\bibnamefont {Zweben}},
  \bibinfo {author} {\bibfnamefont {K.}~\bibnamefont {Hallatschek}}, \bibinfo
  {author} {\bibfnamefont {B.}~\bibnamefont {LaBombard}}, \bibinfo {author}
  {\bibfnamefont {R.}~\bibnamefont {Maqueda}}, \bibinfo {author} {\bibfnamefont
  {B.}~\bibnamefont {Bai}}, \bibinfo {author} {\bibfnamefont {C.}~\bibnamefont
  {Boswell}}, \bibinfo {author} {\bibfnamefont {M.}~\bibnamefont {Greenwald}},
  \bibinfo {author} {\bibfnamefont {D.}~\bibnamefont {Kopon}}, \bibinfo
  {author} {\bibfnamefont {W.}~\bibnamefont {Nevins}},  \emph {et~al.},\
  }\bibfield  {title} {\enquote {\bibinfo {title} {Observations of the
  turbulence in the scrape-off-layer of alcator c-mod and comparisons with
  simulation},}\ }\href@noop {} {\bibfield  {journal} {\bibinfo  {journal}
  {Physics of Plasmas}\ }\textbf {\bibinfo {volume} {10}},\ \bibinfo {pages}
  {1739--1747} (\bibinfo {year} {2003})}\BibitemShut {NoStop}%
\bibitem [{\citenamefont {Myra}\ \emph {et~al.}(2006)\citenamefont {Myra},
  \citenamefont {D’Ippolito}, \citenamefont {Stotler}, \citenamefont
  {Zweben}, \citenamefont {LeBlanc}, \citenamefont {Menard}, \citenamefont
  {Maqueda},\ and\ \citenamefont {Boedo}}]{myra2006blob}%
  \BibitemOpen
  \bibfield  {author} {\bibinfo {author} {\bibfnamefont {J.}~\bibnamefont
  {Myra}}, \bibinfo {author} {\bibfnamefont {D.}~\bibnamefont {D’Ippolito}},
  \bibinfo {author} {\bibfnamefont {D.}~\bibnamefont {Stotler}}, \bibinfo
  {author} {\bibfnamefont {S.}~\bibnamefont {Zweben}}, \bibinfo {author}
  {\bibfnamefont {B.}~\bibnamefont {LeBlanc}}, \bibinfo {author} {\bibfnamefont
  {J.}~\bibnamefont {Menard}}, \bibinfo {author} {\bibfnamefont
  {R.}~\bibnamefont {Maqueda}}, \ and\ \bibinfo {author} {\bibfnamefont
  {J.}~\bibnamefont {Boedo}},\ }\bibfield  {title} {\enquote {\bibinfo {title}
  {Blob birth and transport in the tokamak edge plasma: Analysis of imaging
  data},}\ }\href@noop {} {\bibfield  {journal} {\bibinfo  {journal} {Physics
  of plasmas}\ }\textbf {\bibinfo {volume} {13}},\ \bibinfo {pages} {092509}
  (\bibinfo {year} {2006})}\BibitemShut {NoStop}%
\bibitem [{\citenamefont {Grulke}\ \emph {et~al.}(2006)\citenamefont {Grulke},
  \citenamefont {Terry}, \citenamefont {LaBombard},\ and\ \citenamefont
  {Zweben}}]{grulke2006radially}%
  \BibitemOpen
  \bibfield  {author} {\bibinfo {author} {\bibfnamefont {O.}~\bibnamefont
  {Grulke}}, \bibinfo {author} {\bibfnamefont {J.}~\bibnamefont {Terry}},
  \bibinfo {author} {\bibfnamefont {B.}~\bibnamefont {LaBombard}}, \ and\
  \bibinfo {author} {\bibfnamefont {S.}~\bibnamefont {Zweben}},\ }\bibfield
  {title} {\enquote {\bibinfo {title} {Radially propagating fluctuation
  structures in the scrape-off layer of alcator c-mod},}\ }\href@noop {}
  {\bibfield  {journal} {\bibinfo  {journal} {Physics of Plasmas}\ }\textbf
  {\bibinfo {volume} {13}},\ \bibinfo {pages} {012306} (\bibinfo {year}
  {2006})}\BibitemShut {NoStop}%
\bibitem [{\citenamefont {Zweben}\ \emph {et~al.}(2015)\citenamefont {Zweben},
  \citenamefont {Davis}, \citenamefont {Kaye}, \citenamefont {Myra},
  \citenamefont {Bell}, \citenamefont {LeBlanc}, \citenamefont {Maqueda},
  \citenamefont {Munsat}, \citenamefont {Sabbagh}, \citenamefont {Sechrest}
  \emph {et~al.}}]{zweben2015edge}%
  \BibitemOpen
  \bibfield  {author} {\bibinfo {author} {\bibfnamefont {S.}~\bibnamefont
  {Zweben}}, \bibinfo {author} {\bibfnamefont {W.}~\bibnamefont {Davis}},
  \bibinfo {author} {\bibfnamefont {S.}~\bibnamefont {Kaye}}, \bibinfo {author}
  {\bibfnamefont {J.}~\bibnamefont {Myra}}, \bibinfo {author} {\bibfnamefont
  {R.}~\bibnamefont {Bell}}, \bibinfo {author} {\bibfnamefont {B.}~\bibnamefont
  {LeBlanc}}, \bibinfo {author} {\bibfnamefont {R.}~\bibnamefont {Maqueda}},
  \bibinfo {author} {\bibfnamefont {T.}~\bibnamefont {Munsat}}, \bibinfo
  {author} {\bibfnamefont {S.}~\bibnamefont {Sabbagh}}, \bibinfo {author}
  {\bibfnamefont {Y.}~\bibnamefont {Sechrest}},  \emph {et~al.},\ }\bibfield
  {title} {\enquote {\bibinfo {title} {Edge and sol turbulence and blob
  variations over a large database in nstx},}\ }\href@noop {} {\bibfield
  {journal} {\bibinfo  {journal} {Nuclear Fusion}\ }\textbf {\bibinfo {volume}
  {55}},\ \bibinfo {pages} {093035} (\bibinfo {year} {2015})}\BibitemShut
  {NoStop}%
\bibitem [{\citenamefont {Zweben}\ \emph {et~al.}(2016)\citenamefont {Zweben},
  \citenamefont {Myra}, \citenamefont {Davis}, \citenamefont {D’Ippolito},
  \citenamefont {Gray}, \citenamefont {Kaye}, \citenamefont {LeBlanc},
  \citenamefont {Maqueda}, \citenamefont {Russell}, \citenamefont {Stotler}
  \emph {et~al.}}]{zweben2016blob}%
  \BibitemOpen
  \bibfield  {author} {\bibinfo {author} {\bibfnamefont {S.}~\bibnamefont
  {Zweben}}, \bibinfo {author} {\bibfnamefont {J.}~\bibnamefont {Myra}},
  \bibinfo {author} {\bibfnamefont {W.}~\bibnamefont {Davis}}, \bibinfo
  {author} {\bibfnamefont {D.}~\bibnamefont {D’Ippolito}}, \bibinfo {author}
  {\bibfnamefont {T.}~\bibnamefont {Gray}}, \bibinfo {author} {\bibfnamefont
  {S.}~\bibnamefont {Kaye}}, \bibinfo {author} {\bibfnamefont {B.}~\bibnamefont
  {LeBlanc}}, \bibinfo {author} {\bibfnamefont {R.}~\bibnamefont {Maqueda}},
  \bibinfo {author} {\bibfnamefont {D.}~\bibnamefont {Russell}}, \bibinfo
  {author} {\bibfnamefont {D.}~\bibnamefont {Stotler}},  \emph {et~al.},\
  }\bibfield  {title} {\enquote {\bibinfo {title} {Blob structure and motion in
  the edge and sol of nstx},}\ }\href@noop {} {\bibfield  {journal} {\bibinfo
  {journal} {Plasma Physics and Controlled Fusion}\ }\textbf {\bibinfo {volume}
  {58}},\ \bibinfo {pages} {044007} (\bibinfo {year} {2016})}\BibitemShut
  {NoStop}%
\bibitem [{\citenamefont {D'Ippolito}\ \emph {et~al.}(2004)\citenamefont
  {D'Ippolito}, \citenamefont {Myra}, \citenamefont {Krasheninnikov},
  \citenamefont {Yu},\ and\ \citenamefont {Pigarov}}]{d2004blob}%
  \BibitemOpen
  \bibfield  {author} {\bibinfo {author} {\bibfnamefont {D.}~\bibnamefont
  {D'Ippolito}}, \bibinfo {author} {\bibfnamefont {J.}~\bibnamefont {Myra}},
  \bibinfo {author} {\bibfnamefont {S.}~\bibnamefont {Krasheninnikov}},
  \bibinfo {author} {\bibfnamefont {G.}~\bibnamefont {Yu}}, \ and\ \bibinfo
  {author} {\bibfnamefont {A.~Y.}\ \bibnamefont {Pigarov}},\ }\bibfield
  {title} {\enquote {\bibinfo {title} {Blob transport in the tokamak
  scrape-off-layer},}\ }\href@noop {} {\bibfield  {journal} {\bibinfo
  {journal} {Contributions to Plasma Physics}\ }\textbf {\bibinfo {volume}
  {44}},\ \bibinfo {pages} {205--216} (\bibinfo {year} {2004})}\BibitemShut
  {NoStop}%
\bibitem [{\citenamefont {D’ippolito}, \citenamefont {Myra},\ and\
  \citenamefont {Zweben}(2011)}]{d2011convective}%
  \BibitemOpen
  \bibfield  {author} {\bibinfo {author} {\bibfnamefont {D.}~\bibnamefont
  {D’ippolito}}, \bibinfo {author} {\bibfnamefont {J.}~\bibnamefont {Myra}},
  \ and\ \bibinfo {author} {\bibfnamefont {S.}~\bibnamefont {Zweben}},\
  }\bibfield  {title} {\enquote {\bibinfo {title} {Convective transport by
  intermittent blob-filaments: Comparison of theory and experiment},}\
  }\href@noop {} {\bibfield  {journal} {\bibinfo  {journal} {Physics of
  Plasmas}\ }\textbf {\bibinfo {volume} {18}},\ \bibinfo {pages} {060501}
  (\bibinfo {year} {2011})}\BibitemShut {NoStop}%
\bibitem [{\citenamefont {Krasheninnikov}, \citenamefont {D'ippolito},\ and\
  \citenamefont {Myra}(2008)}]{krasheninnikov2008recent}%
  \BibitemOpen
  \bibfield  {author} {\bibinfo {author} {\bibfnamefont {S.}~\bibnamefont
  {Krasheninnikov}}, \bibinfo {author} {\bibfnamefont {D.}~\bibnamefont
  {D'ippolito}}, \ and\ \bibinfo {author} {\bibfnamefont {J.}~\bibnamefont
  {Myra}},\ }\bibfield  {title} {\enquote {\bibinfo {title} {Recent theoretical
  progress in understanding coherent structures in edge and sol turbulence},}\
  }\href@noop {} {\bibfield  {journal} {\bibinfo  {journal} {Journal of Plasma
  Physics}\ }\textbf {\bibinfo {volume} {74}},\ \bibinfo {pages} {679--717}
  (\bibinfo {year} {2008})}\BibitemShut {NoStop}%
\bibitem [{\citenamefont {Garcia}(2009)}]{garcia2009blob}%
  \BibitemOpen
  \bibfield  {author} {\bibinfo {author} {\bibfnamefont {O.}~\bibnamefont
  {Garcia}},\ }\bibfield  {title} {\enquote {\bibinfo {title} {Blob transport
  in the plasma edge: a review},}\ }\href@noop {} {\bibfield  {journal}
  {\bibinfo  {journal} {Plasma and Fusion Research}\ }\textbf {\bibinfo
  {volume} {4}},\ \bibinfo {pages} {019--019} (\bibinfo {year}
  {2009})}\BibitemShut {NoStop}%
\bibitem [{\citenamefont {Garcia}\ \emph {et~al.}(2004)\citenamefont {Garcia},
  \citenamefont {Naulin}, \citenamefont {Nielsen},\ and\ \citenamefont
  {Rasmussen}}]{garcia2004computations}%
  \BibitemOpen
  \bibfield  {author} {\bibinfo {author} {\bibfnamefont {O.}~\bibnamefont
  {Garcia}}, \bibinfo {author} {\bibfnamefont {V.}~\bibnamefont {Naulin}},
  \bibinfo {author} {\bibfnamefont {A.}~\bibnamefont {Nielsen}}, \ and\
  \bibinfo {author} {\bibfnamefont {J.~J.}\ \bibnamefont {Rasmussen}},\
  }\bibfield  {title} {\enquote {\bibinfo {title} {Computations of intermittent
  transport in scrape-off layer plasmas},}\ }\href@noop {} {\bibfield
  {journal} {\bibinfo  {journal} {Physical review letters}\ }\textbf {\bibinfo
  {volume} {92}},\ \bibinfo {pages} {165003} (\bibinfo {year}
  {2004})}\BibitemShut {NoStop}%
\bibitem [{\citenamefont {Russell}, \citenamefont {Myra},\ and\ \citenamefont
  {D’Ippolito}(2009)}]{russell2009saturation}%
  \BibitemOpen
  \bibfield  {author} {\bibinfo {author} {\bibfnamefont {D.}~\bibnamefont
  {Russell}}, \bibinfo {author} {\bibfnamefont {J.}~\bibnamefont {Myra}}, \
  and\ \bibinfo {author} {\bibfnamefont {D.}~\bibnamefont {D’Ippolito}},\
  }\bibfield  {title} {\enquote {\bibinfo {title} {Saturation mechanisms for
  edge turbulence},}\ }\href@noop {} {\bibfield  {journal} {\bibinfo  {journal}
  {Physics of Plasmas}\ }\textbf {\bibinfo {volume} {16}},\ \bibinfo {pages}
  {122304} (\bibinfo {year} {2009})}\BibitemShut {NoStop}%
\bibitem [{\citenamefont {Militello}\ \emph {et~al.}(2012)\citenamefont
  {Militello}, \citenamefont {Fundamenski}, \citenamefont {Naulin},\ and\
  \citenamefont {Nielsen}}]{militello2012simulations}%
  \BibitemOpen
  \bibfield  {author} {\bibinfo {author} {\bibfnamefont {F.}~\bibnamefont
  {Militello}}, \bibinfo {author} {\bibfnamefont {W.}~\bibnamefont
  {Fundamenski}}, \bibinfo {author} {\bibfnamefont {V.}~\bibnamefont {Naulin}},
  \ and\ \bibinfo {author} {\bibfnamefont {A.~H.}\ \bibnamefont {Nielsen}},\
  }\bibfield  {title} {\enquote {\bibinfo {title} {Simulations of edge and
  scrape off layer turbulence in mega ampere spherical tokamak plasmas},}\
  }\href@noop {} {\bibfield  {journal} {\bibinfo  {journal} {Plasma Physics and
  Controlled Fusion}\ }\textbf {\bibinfo {volume} {54}},\ \bibinfo {pages}
  {095011} (\bibinfo {year} {2012})}\BibitemShut {NoStop}%
\bibitem [{\citenamefont {Militello}, \citenamefont {Naulin},\ and\
  \citenamefont {Nielsen}(2013)}]{militello2013numerical}%
  \BibitemOpen
  \bibfield  {author} {\bibinfo {author} {\bibfnamefont {F.}~\bibnamefont
  {Militello}}, \bibinfo {author} {\bibfnamefont {V.}~\bibnamefont {Naulin}}, \
  and\ \bibinfo {author} {\bibfnamefont {A.~H.}\ \bibnamefont {Nielsen}},\
  }\bibfield  {title} {\enquote {\bibinfo {title} {Numerical scalings of the
  decay lengths in the scrape-off layer},}\ }\href@noop {} {\bibfield
  {journal} {\bibinfo  {journal} {Plasma Physics and Controlled Fusion}\
  }\textbf {\bibinfo {volume} {55}},\ \bibinfo {pages} {074010} (\bibinfo
  {year} {2013})}\BibitemShut {NoStop}%
\bibitem [{\citenamefont {Kube}\ and\ \citenamefont
  {Garcia}(2011)}]{kube2011velocity}%
  \BibitemOpen
  \bibfield  {author} {\bibinfo {author} {\bibfnamefont {R.}~\bibnamefont
  {Kube}}\ and\ \bibinfo {author} {\bibfnamefont {O.~E.}\ \bibnamefont
  {Garcia}},\ }\bibfield  {title} {\enquote {\bibinfo {title} {Velocity scaling
  for filament motion in scrape-off layer plasmas},}\ }\href@noop {} {\bibfield
   {journal} {\bibinfo  {journal} {Physics of Plasmas}\ }\textbf {\bibinfo
  {volume} {18}},\ \bibinfo {pages} {102314} (\bibinfo {year}
  {2011})}\BibitemShut {NoStop}%
\bibitem [{\citenamefont {Kube}, \citenamefont {Garcia},\ and\ \citenamefont
  {Wiesenberger}(2016)}]{kube2016amplitude}%
  \BibitemOpen
  \bibfield  {author} {\bibinfo {author} {\bibfnamefont {R.}~\bibnamefont
  {Kube}}, \bibinfo {author} {\bibfnamefont {O.}~\bibnamefont {Garcia}}, \ and\
  \bibinfo {author} {\bibfnamefont {M.}~\bibnamefont {Wiesenberger}},\
  }\bibfield  {title} {\enquote {\bibinfo {title} {Amplitude and size scaling
  for interchange motions of plasma filaments},}\ }\href@noop {} {\bibfield
  {journal} {\bibinfo  {journal} {Physics of Plasmas}\ }\textbf {\bibinfo
  {volume} {23}},\ \bibinfo {pages} {122302} (\bibinfo {year}
  {2016})}\BibitemShut {NoStop}%
\bibitem [{\citenamefont {Wiesenberger}\ \emph {et~al.}(2017)\citenamefont
  {Wiesenberger}, \citenamefont {Held}, \citenamefont {Kube},\ and\
  \citenamefont {Garcia}}]{wiesenberger2017unified}%
  \BibitemOpen
  \bibfield  {author} {\bibinfo {author} {\bibfnamefont {M.}~\bibnamefont
  {Wiesenberger}}, \bibinfo {author} {\bibfnamefont {M.}~\bibnamefont {Held}},
  \bibinfo {author} {\bibfnamefont {R.}~\bibnamefont {Kube}}, \ and\ \bibinfo
  {author} {\bibfnamefont {O.~E.}\ \bibnamefont {Garcia}},\ }\bibfield  {title}
  {\enquote {\bibinfo {title} {Unified transport scaling laws for plasma blobs
  and depletions},}\ }\href@noop {} {\bibfield  {journal} {\bibinfo  {journal}
  {Physics of Plasmas}\ }\textbf {\bibinfo {volume} {24}},\ \bibinfo {pages}
  {064502} (\bibinfo {year} {2017})}\BibitemShut {NoStop}%
\bibitem [{\citenamefont {Angus}, \citenamefont {Umansky},\ and\ \citenamefont
  {Krasheninnikov}(2012)}]{angus2012effect}%
  \BibitemOpen
  \bibfield  {author} {\bibinfo {author} {\bibfnamefont {J.~R.}\ \bibnamefont
  {Angus}}, \bibinfo {author} {\bibfnamefont {M.~V.}\ \bibnamefont {Umansky}},
  \ and\ \bibinfo {author} {\bibfnamefont {S.~I.}\ \bibnamefont
  {Krasheninnikov}},\ }\bibfield  {title} {\enquote {\bibinfo {title} {Effect
  of drift waves on plasma blob dynamics},}\ }\href@noop {} {\bibfield
  {journal} {\bibinfo  {journal} {Physical review letters}\ }\textbf {\bibinfo
  {volume} {108}},\ \bibinfo {pages} {215002} (\bibinfo {year}
  {2012})}\BibitemShut {NoStop}%
\bibitem [{\citenamefont {Walkden}, \citenamefont {Dudson},\ and\ \citenamefont
  {Fishpool}(2013)}]{walkden2013characterization}%
  \BibitemOpen
  \bibfield  {author} {\bibinfo {author} {\bibfnamefont {N.}~\bibnamefont
  {Walkden}}, \bibinfo {author} {\bibfnamefont {B.}~\bibnamefont {Dudson}}, \
  and\ \bibinfo {author} {\bibfnamefont {G.}~\bibnamefont {Fishpool}},\
  }\bibfield  {title} {\enquote {\bibinfo {title} {Characterization of 3d
  filament dynamics in a mast sol flux tube geometry},}\ }\href@noop {}
  {\bibfield  {journal} {\bibinfo  {journal} {Plasma Physics and Controlled
  Fusion}\ }\textbf {\bibinfo {volume} {55}},\ \bibinfo {pages} {105005}
  (\bibinfo {year} {2013})}\BibitemShut {NoStop}%
\bibitem [{\citenamefont {Ricci}\ \emph {et~al.}(2015)\citenamefont {Ricci},
  \citenamefont {Riva}, \citenamefont {Theiler}, \citenamefont {Fasoli},
  \citenamefont {Furno}, \citenamefont {Halpern},\ and\ \citenamefont
  {Loizu}}]{ricci2015approaching}%
  \BibitemOpen
  \bibfield  {author} {\bibinfo {author} {\bibfnamefont {P.}~\bibnamefont
  {Ricci}}, \bibinfo {author} {\bibfnamefont {F.}~\bibnamefont {Riva}},
  \bibinfo {author} {\bibfnamefont {C.}~\bibnamefont {Theiler}}, \bibinfo
  {author} {\bibfnamefont {A.}~\bibnamefont {Fasoli}}, \bibinfo {author}
  {\bibfnamefont {I.}~\bibnamefont {Furno}}, \bibinfo {author} {\bibfnamefont
  {F.}~\bibnamefont {Halpern}}, \ and\ \bibinfo {author} {\bibfnamefont
  {J.}~\bibnamefont {Loizu}},\ }\bibfield  {title} {\enquote {\bibinfo {title}
  {Approaching the investigation of plasma turbulence through a rigorous
  verification and validation procedure: A practical example},}\ }\href@noop {}
  {\bibfield  {journal} {\bibinfo  {journal} {Physics of Plasmas}\ }\textbf
  {\bibinfo {volume} {22}},\ \bibinfo {pages} {055704} (\bibinfo {year}
  {2015})}\BibitemShut {NoStop}%
\bibitem [{\citenamefont {Tamain}\ \emph {et~al.}(2014)\citenamefont {Tamain},
  \citenamefont {Bufferand}, \citenamefont {Ciraolo}, \citenamefont {Colin},
  \citenamefont {Ghendrih}, \citenamefont {Schwander},\ and\ \citenamefont
  {Serre}}]{tamain20143d}%
  \BibitemOpen
  \bibfield  {author} {\bibinfo {author} {\bibfnamefont {P.}~\bibnamefont
  {Tamain}}, \bibinfo {author} {\bibfnamefont {H.}~\bibnamefont {Bufferand}},
  \bibinfo {author} {\bibfnamefont {G.}~\bibnamefont {Ciraolo}}, \bibinfo
  {author} {\bibfnamefont {C.}~\bibnamefont {Colin}}, \bibinfo {author}
  {\bibfnamefont {P.}~\bibnamefont {Ghendrih}}, \bibinfo {author}
  {\bibfnamefont {F.}~\bibnamefont {Schwander}}, \ and\ \bibinfo {author}
  {\bibfnamefont {E.}~\bibnamefont {Serre}},\ }\bibfield  {title} {\enquote
  {\bibinfo {title} {3d properties of edge turbulent transport in full-torus
  simulations and their impact on poloidal asymmetries},}\ }\href@noop {}
  {\bibfield  {journal} {\bibinfo  {journal} {Contributions to Plasma Physics}\
  }\textbf {\bibinfo {volume} {54}},\ \bibinfo {pages} {555--559} (\bibinfo
  {year} {2014})}\BibitemShut {NoStop}%
\bibitem [{\citenamefont {Easy}\ \emph {et~al.}(2014)\citenamefont {Easy},
  \citenamefont {Militello}, \citenamefont {Omotani}, \citenamefont {Dudson},
  \citenamefont {Havl{\'\i}{\v{c}}kov{\'a}}, \citenamefont {Tamain},
  \citenamefont {Naulin},\ and\ \citenamefont {Nielsen}}]{easy2014three}%
  \BibitemOpen
  \bibfield  {author} {\bibinfo {author} {\bibfnamefont {L.}~\bibnamefont
  {Easy}}, \bibinfo {author} {\bibfnamefont {F.}~\bibnamefont {Militello}},
  \bibinfo {author} {\bibfnamefont {J.}~\bibnamefont {Omotani}}, \bibinfo
  {author} {\bibfnamefont {B.}~\bibnamefont {Dudson}}, \bibinfo {author}
  {\bibfnamefont {E.}~\bibnamefont {Havl{\'\i}{\v{c}}kov{\'a}}}, \bibinfo
  {author} {\bibfnamefont {P.}~\bibnamefont {Tamain}}, \bibinfo {author}
  {\bibfnamefont {V.}~\bibnamefont {Naulin}}, \ and\ \bibinfo {author}
  {\bibfnamefont {A.~H.}\ \bibnamefont {Nielsen}},\ }\bibfield  {title}
  {\enquote {\bibinfo {title} {Three dimensional simulations of plasma
  filaments in the scrape off layer: A comparison with models of reduced
  dimensionality},}\ }\href@noop {} {\bibfield  {journal} {\bibinfo  {journal}
  {Physics of Plasmas}\ }\textbf {\bibinfo {volume} {21}},\ \bibinfo {pages}
  {122515} (\bibinfo {year} {2014})}\BibitemShut {NoStop}%
\bibitem [{\citenamefont {Easy}\ \emph {et~al.}(2016)\citenamefont {Easy},
  \citenamefont {Militello}, \citenamefont {Omotani}, \citenamefont {Walkden},\
  and\ \citenamefont {Dudson}}]{easy2016investigation}%
  \BibitemOpen
  \bibfield  {author} {\bibinfo {author} {\bibfnamefont {L.}~\bibnamefont
  {Easy}}, \bibinfo {author} {\bibfnamefont {F.}~\bibnamefont {Militello}},
  \bibinfo {author} {\bibfnamefont {J.}~\bibnamefont {Omotani}}, \bibinfo
  {author} {\bibfnamefont {N.}~\bibnamefont {Walkden}}, \ and\ \bibinfo
  {author} {\bibfnamefont {B.}~\bibnamefont {Dudson}},\ }\bibfield  {title}
  {\enquote {\bibinfo {title} {Investigation of the effect of resistivity on
  scrape off layer filaments using three-dimensional simulations},}\
  }\href@noop {} {\bibfield  {journal} {\bibinfo  {journal} {Physics of
  Plasmas}\ }\textbf {\bibinfo {volume} {23}},\ \bibinfo {pages} {012512}
  (\bibinfo {year} {2016})}\BibitemShut {NoStop}%
\bibitem [{\citenamefont {Militello}\ \emph {et~al.}(2016)\citenamefont
  {Militello}, \citenamefont {Walkden}, \citenamefont {Farley}, \citenamefont
  {Gracias}, \citenamefont {Olsen}, \citenamefont {Riva}, \citenamefont {Easy},
  \citenamefont {Fedorczak}, \citenamefont {Lupelli}, \citenamefont {Madsen}
  \emph {et~al.}}]{militello2016multi}%
  \BibitemOpen
  \bibfield  {author} {\bibinfo {author} {\bibfnamefont {F.}~\bibnamefont
  {Militello}}, \bibinfo {author} {\bibfnamefont {N.}~\bibnamefont {Walkden}},
  \bibinfo {author} {\bibfnamefont {T.}~\bibnamefont {Farley}}, \bibinfo
  {author} {\bibfnamefont {W.}~\bibnamefont {Gracias}}, \bibinfo {author}
  {\bibfnamefont {J.}~\bibnamefont {Olsen}}, \bibinfo {author} {\bibfnamefont
  {F.}~\bibnamefont {Riva}}, \bibinfo {author} {\bibfnamefont {L.}~\bibnamefont
  {Easy}}, \bibinfo {author} {\bibfnamefont {N.}~\bibnamefont {Fedorczak}},
  \bibinfo {author} {\bibfnamefont {I.}~\bibnamefont {Lupelli}}, \bibinfo
  {author} {\bibfnamefont {J.}~\bibnamefont {Madsen}},  \emph {et~al.},\
  }\bibfield  {title} {\enquote {\bibinfo {title} {Multi-code analysis of
  scrape-off layer filament dynamics in mast},}\ }\href@noop {} {\bibfield
  {journal} {\bibinfo  {journal} {Plasma Physics and Controlled Fusion}\
  }\textbf {\bibinfo {volume} {58}},\ \bibinfo {pages} {105002} (\bibinfo
  {year} {2016})}\BibitemShut {NoStop}%
\bibitem [{\citenamefont {Riva}\ \emph {et~al.}(2016)\citenamefont {Riva},
  \citenamefont {Colin}, \citenamefont {Denis}, \citenamefont {Easy},
  \citenamefont {Furno}, \citenamefont {Madsen}, \citenamefont {Militello},
  \citenamefont {Naulin}, \citenamefont {Nielsen}, \citenamefont {Olsen} \emph
  {et~al.}}]{riva2016blob}%
  \BibitemOpen
  \bibfield  {author} {\bibinfo {author} {\bibfnamefont {F.}~\bibnamefont
  {Riva}}, \bibinfo {author} {\bibfnamefont {C.}~\bibnamefont {Colin}},
  \bibinfo {author} {\bibfnamefont {J.}~\bibnamefont {Denis}}, \bibinfo
  {author} {\bibfnamefont {L.}~\bibnamefont {Easy}}, \bibinfo {author}
  {\bibfnamefont {I.}~\bibnamefont {Furno}}, \bibinfo {author} {\bibfnamefont
  {J.}~\bibnamefont {Madsen}}, \bibinfo {author} {\bibfnamefont
  {F.}~\bibnamefont {Militello}}, \bibinfo {author} {\bibfnamefont
  {V.}~\bibnamefont {Naulin}}, \bibinfo {author} {\bibfnamefont {A.~H.}\
  \bibnamefont {Nielsen}}, \bibinfo {author} {\bibfnamefont {J.~M.~B.}\
  \bibnamefont {Olsen}},  \emph {et~al.},\ }\bibfield  {title} {\enquote
  {\bibinfo {title} {Blob dynamics in the torpex experiment: a multi-code
  validation},}\ }\href@noop {} {\bibfield  {journal} {\bibinfo  {journal}
  {Plasma Physics and Controlled Fusion}\ }\textbf {\bibinfo {volume} {58}},\
  \bibinfo {pages} {044005} (\bibinfo {year} {2016})}\BibitemShut {NoStop}%
\bibitem [{\citenamefont {Riva}\ \emph {et~al.}(2019)\citenamefont {Riva},
  \citenamefont {Militello}, \citenamefont {Elmore}, \citenamefont {Omotani},
  \citenamefont {Dudson},\ and\ \citenamefont {Walkden}}]{riva2019three}%
  \BibitemOpen
  \bibfield  {author} {\bibinfo {author} {\bibfnamefont {F.}~\bibnamefont
  {Riva}}, \bibinfo {author} {\bibfnamefont {F.}~\bibnamefont {Militello}},
  \bibinfo {author} {\bibfnamefont {S.}~\bibnamefont {Elmore}}, \bibinfo
  {author} {\bibfnamefont {J.~T.}\ \bibnamefont {Omotani}}, \bibinfo {author}
  {\bibfnamefont {B.~D.}\ \bibnamefont {Dudson}}, \ and\ \bibinfo {author}
  {\bibfnamefont {N.}~\bibnamefont {Walkden}},\ }\bibfield  {title} {\enquote
  {\bibinfo {title} {Three-dimensional plasma edge turbulence simulations of
  mast and comparison with experimental measurements},}\ }\href@noop {}
  {\bibfield  {journal} {\bibinfo  {journal} {Plasma Physics and Controlled
  Fusion}\ } (\bibinfo {year} {2019})}\BibitemShut {NoStop}%
\bibitem [{\citenamefont {Wiesenberger}, \citenamefont {Madsen},\ and\
  \citenamefont {Kendl}(2014)}]{wiesenberger2014radial}%
  \BibitemOpen
  \bibfield  {author} {\bibinfo {author} {\bibfnamefont {M.}~\bibnamefont
  {Wiesenberger}}, \bibinfo {author} {\bibfnamefont {J.}~\bibnamefont
  {Madsen}}, \ and\ \bibinfo {author} {\bibfnamefont {A.}~\bibnamefont
  {Kendl}},\ }\bibfield  {title} {\enquote {\bibinfo {title} {Radial convection
  of finite ion temperature, high amplitude plasma blobs},}\ }\href@noop {}
  {\bibfield  {journal} {\bibinfo  {journal} {Physics of Plasmas}\ }\textbf
  {\bibinfo {volume} {21}},\ \bibinfo {pages} {092301} (\bibinfo {year}
  {2014})}\BibitemShut {NoStop}%
\bibitem [{\citenamefont {Lee}\ \emph {et~al.}(2015)\citenamefont {Lee},
  \citenamefont {Angus}, \citenamefont {Umansky},\ and\ \citenamefont
  {Krasheninnikov}}]{lee2015electromagnetic}%
  \BibitemOpen
  \bibfield  {author} {\bibinfo {author} {\bibfnamefont {W.}~\bibnamefont
  {Lee}}, \bibinfo {author} {\bibfnamefont {J.~R.}\ \bibnamefont {Angus}},
  \bibinfo {author} {\bibfnamefont {M.~V.}\ \bibnamefont {Umansky}}, \ and\
  \bibinfo {author} {\bibfnamefont {S.~I.}\ \bibnamefont {Krasheninnikov}},\
  }\bibfield  {title} {\enquote {\bibinfo {title} {Electromagnetic effects on
  plasma blob-filament transport},}\ }\href@noop {} {\bibfield  {journal}
  {\bibinfo  {journal} {Journal of Nuclear Materials}\ }\textbf {\bibinfo
  {volume} {463}},\ \bibinfo {pages} {765--768} (\bibinfo {year}
  {2015})}\BibitemShut {NoStop}%
\bibitem [{\citenamefont {Angus}, \citenamefont {Krasheninnikov},\ and\
  \citenamefont {Umansky}(2012)}]{angus2012effects}%
  \BibitemOpen
  \bibfield  {author} {\bibinfo {author} {\bibfnamefont {J.~R.}\ \bibnamefont
  {Angus}}, \bibinfo {author} {\bibfnamefont {S.~I.}\ \bibnamefont
  {Krasheninnikov}}, \ and\ \bibinfo {author} {\bibfnamefont {M.~V.}\
  \bibnamefont {Umansky}},\ }\bibfield  {title} {\enquote {\bibinfo {title}
  {Effects of parallel electron dynamics on plasma blob transport},}\
  }\href@noop {} {\bibfield  {journal} {\bibinfo  {journal} {Physics of
  Plasmas}\ }\textbf {\bibinfo {volume} {19}},\ \bibinfo {pages} {082312}
  (\bibinfo {year} {2012})}\BibitemShut {NoStop}%
\bibitem [{\citenamefont {Myra}, \citenamefont {Russell},\ and\ \citenamefont
  {D’Ippolito}(2006)}]{myra2006collisionality}%
  \BibitemOpen
  \bibfield  {author} {\bibinfo {author} {\bibfnamefont {J.}~\bibnamefont
  {Myra}}, \bibinfo {author} {\bibfnamefont {D.}~\bibnamefont {Russell}}, \
  and\ \bibinfo {author} {\bibfnamefont {D.}~\bibnamefont {D’Ippolito}},\
  }\bibfield  {title} {\enquote {\bibinfo {title} {Collisionality and magnetic
  geometry effects on tokamak edge turbulent transport. i. a two-region model
  with application to blobs},}\ }\href@noop {} {\bibfield  {journal} {\bibinfo
  {journal} {Physics of plasmas}\ }\textbf {\bibinfo {volume} {13}},\ \bibinfo
  {pages} {112502} (\bibinfo {year} {2006})}\BibitemShut {NoStop}%
\bibitem [{\citenamefont {Russell}, \citenamefont {Myra},\ and\ \citenamefont
  {D’Ippolito}(2007)}]{russell2007collisionality}%
  \BibitemOpen
  \bibfield  {author} {\bibinfo {author} {\bibfnamefont {D.}~\bibnamefont
  {Russell}}, \bibinfo {author} {\bibfnamefont {J.}~\bibnamefont {Myra}}, \
  and\ \bibinfo {author} {\bibfnamefont {D.}~\bibnamefont {D’Ippolito}},\
  }\bibfield  {title} {\enquote {\bibinfo {title} {Collisionality and magnetic
  geometry effects on tokamak edge turbulent transport. ii. many-blob
  turbulence in the two-region model},}\ }\href@noop {} {\bibfield  {journal}
  {\bibinfo  {journal} {Physics of Plasmas}\ }\textbf {\bibinfo {volume}
  {14}},\ \bibinfo {pages} {102307} (\bibinfo {year} {2007})}\BibitemShut
  {NoStop}%
\bibitem [{\citenamefont {Militello}\ \emph {et~al.}(2017)\citenamefont
  {Militello}, \citenamefont {Dudson}, \citenamefont {Easy}, \citenamefont
  {Kirk},\ and\ \citenamefont {Naylor}}]{militello2017interaction}%
  \BibitemOpen
  \bibfield  {author} {\bibinfo {author} {\bibfnamefont {F.}~\bibnamefont
  {Militello}}, \bibinfo {author} {\bibfnamefont {B.}~\bibnamefont {Dudson}},
  \bibinfo {author} {\bibfnamefont {L.}~\bibnamefont {Easy}}, \bibinfo {author}
  {\bibfnamefont {A.}~\bibnamefont {Kirk}}, \ and\ \bibinfo {author}
  {\bibfnamefont {P.}~\bibnamefont {Naylor}},\ }\bibfield  {title} {\enquote
  {\bibinfo {title} {On the interaction of scrape off layer filaments},}\
  }\href@noop {} {\bibfield  {journal} {\bibinfo  {journal} {Plasma Physics and
  Controlled Fusion}\ }\textbf {\bibinfo {volume} {59}},\ \bibinfo {pages}
  {125013} (\bibinfo {year} {2017})}\BibitemShut {NoStop}%
\bibitem [{\citenamefont {Nespoli}\ \emph {et~al.}(2017)\citenamefont
  {Nespoli}, \citenamefont {Furno}, \citenamefont {Labit}, \citenamefont
  {Ricci}, \citenamefont {Avino}, \citenamefont {Halpern}, \citenamefont
  {Musil},\ and\ \citenamefont {Riva}}]{nespoli2017blob}%
  \BibitemOpen
  \bibfield  {author} {\bibinfo {author} {\bibfnamefont {F.}~\bibnamefont
  {Nespoli}}, \bibinfo {author} {\bibfnamefont {I.}~\bibnamefont {Furno}},
  \bibinfo {author} {\bibfnamefont {B.}~\bibnamefont {Labit}}, \bibinfo
  {author} {\bibfnamefont {P.}~\bibnamefont {Ricci}}, \bibinfo {author}
  {\bibfnamefont {F.}~\bibnamefont {Avino}}, \bibinfo {author} {\bibfnamefont
  {F.}~\bibnamefont {Halpern}}, \bibinfo {author} {\bibfnamefont
  {F.}~\bibnamefont {Musil}}, \ and\ \bibinfo {author} {\bibfnamefont
  {F.}~\bibnamefont {Riva}},\ }\bibfield  {title} {\enquote {\bibinfo {title}
  {Blob properties in full-turbulence simulations of the tcv scrape-off
  layer},}\ }\href@noop {} {\bibfield  {journal} {\bibinfo  {journal} {Plasma
  Physics and Controlled Fusion}\ }\textbf {\bibinfo {volume} {59}},\ \bibinfo
  {pages} {055009} (\bibinfo {year} {2017})}\BibitemShut {NoStop}%
\bibitem [{\citenamefont {Nespoli}\ \emph {et~al.}(2019)\citenamefont
  {Nespoli}, \citenamefont {Tamain}, \citenamefont {Fedorczak}, \citenamefont
  {Ciraolo}, \citenamefont {Galassi}, \citenamefont {Tatali}, \citenamefont
  {Serre}, \citenamefont {Marandet}, \citenamefont {Bufferand},\ and\
  \citenamefont {Ghendrih}}]{nespoli20193d}%
  \BibitemOpen
  \bibfield  {author} {\bibinfo {author} {\bibfnamefont {F.}~\bibnamefont
  {Nespoli}}, \bibinfo {author} {\bibfnamefont {P.}~\bibnamefont {Tamain}},
  \bibinfo {author} {\bibfnamefont {N.}~\bibnamefont {Fedorczak}}, \bibinfo
  {author} {\bibfnamefont {G.}~\bibnamefont {Ciraolo}}, \bibinfo {author}
  {\bibfnamefont {D.}~\bibnamefont {Galassi}}, \bibinfo {author} {\bibfnamefont
  {R.}~\bibnamefont {Tatali}}, \bibinfo {author} {\bibfnamefont
  {E.}~\bibnamefont {Serre}}, \bibinfo {author} {\bibfnamefont
  {Y.}~\bibnamefont {Marandet}}, \bibinfo {author} {\bibfnamefont
  {H.}~\bibnamefont {Bufferand}}, \ and\ \bibinfo {author} {\bibfnamefont
  {P.}~\bibnamefont {Ghendrih}},\ }\bibfield  {title} {\enquote {\bibinfo
  {title} {3d structure and dynamics of filaments in turbulence simulations of
  west diverted plasmas},}\ }\href@noop {} {\bibfield  {journal} {\bibinfo
  {journal} {Nuclear Fusion}\ }\textbf {\bibinfo {volume} {59}},\ \bibinfo
  {pages} {096006} (\bibinfo {year} {2019})}\BibitemShut {NoStop}%
\bibitem [{\citenamefont {Paruta}\ \emph {et~al.}(2019)\citenamefont {Paruta},
  \citenamefont {Beadle}, \citenamefont {Ricci},\ and\ \citenamefont
  {Theiler}}]{paruta2019blob}%
  \BibitemOpen
  \bibfield  {author} {\bibinfo {author} {\bibfnamefont {P.}~\bibnamefont
  {Paruta}}, \bibinfo {author} {\bibfnamefont {C.}~\bibnamefont {Beadle}},
  \bibinfo {author} {\bibfnamefont {P.}~\bibnamefont {Ricci}}, \ and\ \bibinfo
  {author} {\bibfnamefont {C.}~\bibnamefont {Theiler}},\ }\bibfield  {title}
  {\enquote {\bibinfo {title} {Blob velocity scaling in diverted tokamaks: A
  comparison between theory and simulation},}\ }\href@noop {} {\bibfield
  {journal} {\bibinfo  {journal} {Physics of Plasmas}\ }\textbf {\bibinfo
  {volume} {26}},\ \bibinfo {pages} {032302} (\bibinfo {year}
  {2019})}\BibitemShut {NoStop}%
\bibitem [{\citenamefont {Ko{\v{c}}an}\ \emph {et~al.}(2007)\citenamefont
  {Ko{\v{c}}an}, \citenamefont {P{\'a}nek}, \citenamefont {St{\"o}ckel},
  \citenamefont {Hron}, \citenamefont {Gunn},\ and\ \citenamefont
  {Dejarnac}}]{kovcan2007ion}%
  \BibitemOpen
  \bibfield  {author} {\bibinfo {author} {\bibfnamefont {M.}~\bibnamefont
  {Ko{\v{c}}an}}, \bibinfo {author} {\bibfnamefont {R.}~\bibnamefont
  {P{\'a}nek}}, \bibinfo {author} {\bibfnamefont {J.}~\bibnamefont
  {St{\"o}ckel}}, \bibinfo {author} {\bibfnamefont {M.}~\bibnamefont {Hron}},
  \bibinfo {author} {\bibfnamefont {J.}~\bibnamefont {Gunn}}, \ and\ \bibinfo
  {author} {\bibfnamefont {R.}~\bibnamefont {Dejarnac}},\ }\bibfield  {title}
  {\enquote {\bibinfo {title} {Ion temperature measurements in the tokamak
  scrape-off layer},}\ }\href@noop {} {\bibfield  {journal} {\bibinfo
  {journal} {Journal of nuclear materials}\ }\textbf {\bibinfo {volume}
  {363}},\ \bibinfo {pages} {1436--1440} (\bibinfo {year} {2007})}\BibitemShut
  {NoStop}%
\bibitem [{\citenamefont {Kocan}\ \emph {et~al.}(2012)\citenamefont {Kocan},
  \citenamefont {Gennrich}, \citenamefont {Kendl}, \citenamefont {M{\"u}ller},\
  and\ \citenamefont {Team}}]{kocan2012ion}%
  \BibitemOpen
  \bibfield  {author} {\bibinfo {author} {\bibfnamefont {M.}~\bibnamefont
  {Kocan}}, \bibinfo {author} {\bibfnamefont {F.}~\bibnamefont {Gennrich}},
  \bibinfo {author} {\bibfnamefont {A.}~\bibnamefont {Kendl}}, \bibinfo
  {author} {\bibfnamefont {H.}~\bibnamefont {M{\"u}ller}}, \ and\ \bibinfo
  {author} {\bibfnamefont {A.~U.}\ \bibnamefont {Team}},\ }\bibfield  {title}
  {\enquote {\bibinfo {title} {Ion temperature fluctuations in the asdex
  upgrade scrape-off layer},}\ }\href@noop {} {\bibfield  {journal} {\bibinfo
  {journal} {Plasma Physics and Controlled Fusion}\ }\textbf {\bibinfo {volume}
  {54}} (\bibinfo {year} {2012})}\BibitemShut {NoStop}%
\bibitem [{\citenamefont {Garcia}\ \emph {et~al.}(2005)\citenamefont {Garcia},
  \citenamefont {Horacek}, \citenamefont {Pitts}, \citenamefont {Nielsen},
  \citenamefont {Fundamenski}, \citenamefont {Graves}, \citenamefont {Naulin},\
  and\ \citenamefont {Rasmussen}}]{garcia2005interchange}%
  \BibitemOpen
  \bibfield  {author} {\bibinfo {author} {\bibfnamefont {O.}~\bibnamefont
  {Garcia}}, \bibinfo {author} {\bibfnamefont {J.}~\bibnamefont {Horacek}},
  \bibinfo {author} {\bibfnamefont {R.}~\bibnamefont {Pitts}}, \bibinfo
  {author} {\bibfnamefont {A.}~\bibnamefont {Nielsen}}, \bibinfo {author}
  {\bibfnamefont {W.}~\bibnamefont {Fundamenski}}, \bibinfo {author}
  {\bibfnamefont {J.}~\bibnamefont {Graves}}, \bibinfo {author} {\bibfnamefont
  {V.}~\bibnamefont {Naulin}}, \ and\ \bibinfo {author} {\bibfnamefont {J.~J.}\
  \bibnamefont {Rasmussen}},\ }\bibfield  {title} {\enquote {\bibinfo {title}
  {Interchange turbulence in the tcv scrape-off layer},}\ }\href@noop {}
  {\bibfield  {journal} {\bibinfo  {journal} {Plasma physics and controlled
  fusion}\ }\textbf {\bibinfo {volume} {48}},\ \bibinfo {pages} {L1} (\bibinfo
  {year} {2005})}\BibitemShut {NoStop}%
\bibitem [{\citenamefont {Dudson}\ \emph {et~al.}(2009)\citenamefont {Dudson},
  \citenamefont {Umansky}, \citenamefont {Xu}, \citenamefont {Snyder},\ and\
  \citenamefont {Wilson}}]{dudson2009bout++}%
  \BibitemOpen
  \bibfield  {author} {\bibinfo {author} {\bibfnamefont {B.}~\bibnamefont
  {Dudson}}, \bibinfo {author} {\bibfnamefont {M.}~\bibnamefont {Umansky}},
  \bibinfo {author} {\bibfnamefont {X.}~\bibnamefont {Xu}}, \bibinfo {author}
  {\bibfnamefont {P.}~\bibnamefont {Snyder}}, \ and\ \bibinfo {author}
  {\bibfnamefont {H.}~\bibnamefont {Wilson}},\ }\bibfield  {title} {\enquote
  {\bibinfo {title} {Bout++: A framework for parallel plasma fluid
  simulations},}\ }\href@noop {} {\bibfield  {journal} {\bibinfo  {journal}
  {Computer Physics Communications}\ }\textbf {\bibinfo {volume} {180}},\
  \bibinfo {pages} {1467--1480} (\bibinfo {year} {2009})}\BibitemShut {NoStop}%
\bibitem [{\citenamefont {Dudson}\ \emph {et~al.}(2019)\citenamefont {Dudson},
  \citenamefont {Hill}, \citenamefont {Dickinson}, \citenamefont {Parker},
  \citenamefont {Allen}, \citenamefont {Breyiannia}, \citenamefont {Brown},
  \citenamefont {Easy}, \citenamefont {Farley}, \citenamefont {Friedman},
  \citenamefont {Grinaker}, \citenamefont {Izacard}, \citenamefont {Joseph},
  \citenamefont {Kim}, \citenamefont {Leconte}, \citenamefont {Leddy},
  \citenamefont {Løiten}, \citenamefont {Ma}, \citenamefont {Madsen},
  \citenamefont {Meyerson}, \citenamefont {Naylor}, \citenamefont {Myers},
  \citenamefont {Omotani}, \citenamefont {Rhee}, \citenamefont {Sauppe},
  \citenamefont {Savage}, \citenamefont {Seto}, \citenamefont {Schwörer},
  \citenamefont {Shanahan}, \citenamefont {Thomas}, \citenamefont {Tiwari},
  \citenamefont {Umansky}, \citenamefont {Walkden}, \citenamefont {Wang},
  \citenamefont {Wang}, \citenamefont {Xi}, \citenamefont {Xia}, \citenamefont
  {Xu}, \citenamefont {Zhang}, \citenamefont {Bokshi}, \citenamefont
  {Muhammed},\ and\ \citenamefont {Estarellas}}]{dudson_ben_2019_3518905}%
  \BibitemOpen
  \bibfield  {author} {\bibinfo {author} {\bibfnamefont {B.}~\bibnamefont
  {Dudson}}, \bibinfo {author} {\bibfnamefont {P.}~\bibnamefont {Hill}},
  \bibinfo {author} {\bibfnamefont {D.}~\bibnamefont {Dickinson}}, \bibinfo
  {author} {\bibfnamefont {J.}~\bibnamefont {Parker}}, \bibinfo {author}
  {\bibfnamefont {A.}~\bibnamefont {Allen}}, \bibinfo {author} {\bibfnamefont
  {G.}~\bibnamefont {Breyiannia}}, \bibinfo {author} {\bibfnamefont
  {J.}~\bibnamefont {Brown}}, \bibinfo {author} {\bibfnamefont
  {L.}~\bibnamefont {Easy}}, \bibinfo {author} {\bibfnamefont {S.}~\bibnamefont
  {Farley}}, \bibinfo {author} {\bibfnamefont {B.}~\bibnamefont {Friedman}},
  \bibinfo {author} {\bibfnamefont {E.}~\bibnamefont {Grinaker}}, \bibinfo
  {author} {\bibfnamefont {O.}~\bibnamefont {Izacard}}, \bibinfo {author}
  {\bibfnamefont {I.}~\bibnamefont {Joseph}}, \bibinfo {author} {\bibfnamefont
  {M.}~\bibnamefont {Kim}}, \bibinfo {author} {\bibfnamefont {M.}~\bibnamefont
  {Leconte}}, \bibinfo {author} {\bibfnamefont {J.}~\bibnamefont {Leddy}},
  \bibinfo {author} {\bibfnamefont {M.}~\bibnamefont {Løiten}}, \bibinfo
  {author} {\bibfnamefont {C.}~\bibnamefont {Ma}}, \bibinfo {author}
  {\bibfnamefont {J.}~\bibnamefont {Madsen}}, \bibinfo {author} {\bibfnamefont
  {D.}~\bibnamefont {Meyerson}}, \bibinfo {author} {\bibfnamefont
  {P.}~\bibnamefont {Naylor}}, \bibinfo {author} {\bibfnamefont
  {S.}~\bibnamefont {Myers}}, \bibinfo {author} {\bibfnamefont
  {J.}~\bibnamefont {Omotani}}, \bibinfo {author} {\bibfnamefont
  {T.}~\bibnamefont {Rhee}}, \bibinfo {author} {\bibfnamefont {J.}~\bibnamefont
  {Sauppe}}, \bibinfo {author} {\bibfnamefont {K.}~\bibnamefont {Savage}},
  \bibinfo {author} {\bibfnamefont {H.}~\bibnamefont {Seto}}, \bibinfo {author}
  {\bibfnamefont {D.}~\bibnamefont {Schwörer}}, \bibinfo {author}
  {\bibfnamefont {B.}~\bibnamefont {Shanahan}}, \bibinfo {author}
  {\bibfnamefont {M.}~\bibnamefont {Thomas}}, \bibinfo {author} {\bibfnamefont
  {S.}~\bibnamefont {Tiwari}}, \bibinfo {author} {\bibfnamefont
  {M.}~\bibnamefont {Umansky}}, \bibinfo {author} {\bibfnamefont
  {N.}~\bibnamefont {Walkden}}, \bibinfo {author} {\bibfnamefont
  {L.}~\bibnamefont {Wang}}, \bibinfo {author} {\bibfnamefont {Z.}~\bibnamefont
  {Wang}}, \bibinfo {author} {\bibfnamefont {P.}~\bibnamefont {Xi}}, \bibinfo
  {author} {\bibfnamefont {T.}~\bibnamefont {Xia}}, \bibinfo {author}
  {\bibfnamefont {X.}~\bibnamefont {Xu}}, \bibinfo {author} {\bibfnamefont
  {H.}~\bibnamefont {Zhang}}, \bibinfo {author} {\bibfnamefont
  {A.}~\bibnamefont {Bokshi}}, \bibinfo {author} {\bibfnamefont
  {H.}~\bibnamefont {Muhammed}}, \ and\ \bibinfo {author} {\bibfnamefont
  {M.}~\bibnamefont {Estarellas}},\ }\href {\doibase 10.5281/zenodo.3518905}
  {\enquote {\bibinfo {title} {Bout++ v4.3.0},}\ } (\bibinfo {year}
  {2019})\BibitemShut {NoStop}%
\bibitem [{\citenamefont {Byrne}\ and\ \citenamefont
  {Hindmarsh}(1999)}]{byrne1999pvode}%
  \BibitemOpen
  \bibfield  {author} {\bibinfo {author} {\bibfnamefont {G.~D.}\ \bibnamefont
  {Byrne}}\ and\ \bibinfo {author} {\bibfnamefont {A.~C.}\ \bibnamefont
  {Hindmarsh}},\ }\bibfield  {title} {\enquote {\bibinfo {title} {Pvode, an ode
  solver for parallel computers},}\ }\href@noop {} {\bibfield  {journal}
  {\bibinfo  {journal} {The International Journal of High Performance Computing
  Applications}\ }\textbf {\bibinfo {volume} {13}},\ \bibinfo {pages}
  {354--365} (\bibinfo {year} {1999})}\BibitemShut {NoStop}%
\bibitem [{\citenamefont {Hoyer}\ and\ \citenamefont
  {Hamman}(2017)}]{hoyer2017xarray}%
  \BibitemOpen
  \bibfield  {author} {\bibinfo {author} {\bibfnamefont {S.}~\bibnamefont
  {Hoyer}}\ and\ \bibinfo {author} {\bibfnamefont {J.}~\bibnamefont {Hamman}},\
  }\bibfield  {title} {\enquote {\bibinfo {title} {xarray: Nd labeled arrays
  and datasets in python},}\ }\href@noop {} {\bibfield  {journal} {\bibinfo
  {journal} {Journal of Open Research Software}\ }\textbf {\bibinfo {volume}
  {5}} (\bibinfo {year} {2017})}\BibitemShut {NoStop}%
\bibitem [{\citenamefont {Garcia}\ \emph {et~al.}(2018)\citenamefont {Garcia},
  \citenamefont {Kube}, \citenamefont {Theodorsen}, \citenamefont {LaBombard},\
  and\ \citenamefont {Terry}}]{garcia2018intermittent}%
  \BibitemOpen
  \bibfield  {author} {\bibinfo {author} {\bibfnamefont {O.~E.}\ \bibnamefont
  {Garcia}}, \bibinfo {author} {\bibfnamefont {R.}~\bibnamefont {Kube}},
  \bibinfo {author} {\bibfnamefont {A.}~\bibnamefont {Theodorsen}}, \bibinfo
  {author} {\bibfnamefont {B.}~\bibnamefont {LaBombard}}, \ and\ \bibinfo
  {author} {\bibfnamefont {J.}~\bibnamefont {Terry}},\ }\bibfield  {title}
  {\enquote {\bibinfo {title} {Intermittent fluctuations in the alcator c-mod
  scrape-off layer for ohmic and high confinement mode plasmas},}\ }\href@noop
  {} {\bibfield  {journal} {\bibinfo  {journal} {Physics of Plasmas}\ }\textbf
  {\bibinfo {volume} {25}},\ \bibinfo {pages} {056103} (\bibinfo {year}
  {2018})}\BibitemShut {NoStop}%
\bibitem [{\citenamefont {Kube}\ \emph {et~al.}(2019)\citenamefont {Kube},
  \citenamefont {Garcia}, \citenamefont {Theodorsen}, \citenamefont {Kuang},
  \citenamefont {LaBombard}, \citenamefont {Terry},\ and\ \citenamefont
  {Brunner}}]{kube2019statistical}%
  \BibitemOpen
  \bibfield  {author} {\bibinfo {author} {\bibfnamefont {R.}~\bibnamefont
  {Kube}}, \bibinfo {author} {\bibfnamefont {O.~E.}\ \bibnamefont {Garcia}},
  \bibinfo {author} {\bibfnamefont {A.}~\bibnamefont {Theodorsen}}, \bibinfo
  {author} {\bibfnamefont {A.}~\bibnamefont {Kuang}}, \bibinfo {author}
  {\bibfnamefont {B.}~\bibnamefont {LaBombard}}, \bibinfo {author}
  {\bibfnamefont {J.~L.}\ \bibnamefont {Terry}}, \ and\ \bibinfo {author}
  {\bibfnamefont {D.}~\bibnamefont {Brunner}},\ }\bibfield  {title} {\enquote
  {\bibinfo {title} {Statistical properties of the plasma fluctuations and
  turbulent cross-field fluxes in the outboard mid-plane scrape-off layer of
  alcator c-mod},}\ }\href@noop {} {\bibfield  {journal} {\bibinfo  {journal}
  {Nuclear Materials and Energy}\ }\textbf {\bibinfo {volume} {18}},\ \bibinfo
  {pages} {193--200} (\bibinfo {year} {2019})}\BibitemShut {NoStop}%
\bibitem [{\citenamefont {Theodorsen}\ \emph {et~al.}(2016)\citenamefont
  {Theodorsen}, \citenamefont {Garcia}, \citenamefont {Horacek}, \citenamefont
  {Kube},\ and\ \citenamefont {Pitts}}]{theodorsen2016scrape}%
  \BibitemOpen
  \bibfield  {author} {\bibinfo {author} {\bibfnamefont {A.}~\bibnamefont
  {Theodorsen}}, \bibinfo {author} {\bibfnamefont {O.~E.}\ \bibnamefont
  {Garcia}}, \bibinfo {author} {\bibfnamefont {J.}~\bibnamefont {Horacek}},
  \bibinfo {author} {\bibfnamefont {R.}~\bibnamefont {Kube}}, \ and\ \bibinfo
  {author} {\bibfnamefont {R.}~\bibnamefont {Pitts}},\ }\bibfield  {title}
  {\enquote {\bibinfo {title} {Scrape-off layer turbulence in tcv: Evidence in
  support of stochastic modelling},}\ }\href@noop {} {\bibfield  {journal}
  {\bibinfo  {journal} {Plasma Physics and Controlled Fusion}\ }\textbf
  {\bibinfo {volume} {58}},\ \bibinfo {pages} {044006} (\bibinfo {year}
  {2016})}\BibitemShut {NoStop}%
\bibitem [{\citenamefont {Kube}\ \emph {et~al.}(2018)\citenamefont {Kube},
  \citenamefont {Garcia}, \citenamefont {Theodorsen}, \citenamefont {Brunner},
  \citenamefont {Kuang}, \citenamefont {LaBombard},\ and\ \citenamefont
  {Terry}}]{kube2018intermittent}%
  \BibitemOpen
  \bibfield  {author} {\bibinfo {author} {\bibfnamefont {R.}~\bibnamefont
  {Kube}}, \bibinfo {author} {\bibfnamefont {O.~E.}\ \bibnamefont {Garcia}},
  \bibinfo {author} {\bibfnamefont {A.}~\bibnamefont {Theodorsen}}, \bibinfo
  {author} {\bibfnamefont {D.}~\bibnamefont {Brunner}}, \bibinfo {author}
  {\bibfnamefont {A.}~\bibnamefont {Kuang}}, \bibinfo {author} {\bibfnamefont
  {B.}~\bibnamefont {LaBombard}}, \ and\ \bibinfo {author} {\bibfnamefont
  {J.~L.}\ \bibnamefont {Terry}},\ }\bibfield  {title} {\enquote {\bibinfo
  {title} {Intermittent electron density and temperature fluctuations and
  associated fluxes in the alcator c-mod scrape-off layer},}\ }\href@noop {}
  {\bibfield  {journal} {\bibinfo  {journal} {Plasma Physics and Controlled
  Fusion}\ }\textbf {\bibinfo {volume} {60}},\ \bibinfo {pages} {065002}
  (\bibinfo {year} {2018})}\BibitemShut {NoStop}%
\bibitem [{\citenamefont {Garcia}\ \emph {et~al.}(2017)\citenamefont {Garcia},
  \citenamefont {Kube}, \citenamefont {Theodorsen}, \citenamefont {Bak},
  \citenamefont {Hong}, \citenamefont {Kim}, \citenamefont {Pitts} \emph
  {et~al.}}]{garcia2017sol}%
  \BibitemOpen
  \bibfield  {author} {\bibinfo {author} {\bibfnamefont {O.}~\bibnamefont
  {Garcia}}, \bibinfo {author} {\bibfnamefont {R.}~\bibnamefont {Kube}},
  \bibinfo {author} {\bibfnamefont {A.}~\bibnamefont {Theodorsen}}, \bibinfo
  {author} {\bibfnamefont {J.-G.}\ \bibnamefont {Bak}}, \bibinfo {author}
  {\bibfnamefont {S.-H.}\ \bibnamefont {Hong}}, \bibinfo {author}
  {\bibfnamefont {H.-S.}\ \bibnamefont {Kim}}, \bibinfo {author} {\bibfnamefont
  {R.}~\bibnamefont {Pitts}},  \emph {et~al.},\ }\bibfield  {title} {\enquote
  {\bibinfo {title} {Sol width and intermittent fluctuations in kstar},}\
  }\href@noop {} {\bibfield  {journal} {\bibinfo  {journal} {Nuclear Materials
  and Energy}\ }\textbf {\bibinfo {volume} {12}},\ \bibinfo {pages} {36--43}
  (\bibinfo {year} {2017})}\BibitemShut {NoStop}%
\end{thebibliography}%

\end{document}